\documentclass[12pt]{article}

\global\arraycolsep=1pt
\usepackage{amssymb}
\usepackage{amsmath}
\usepackage {pstricks}
\usepackage{graphicx,color}
\usepackage{mathrsfs}

\usepackage[cmbtt]{bold-extra}

\usepackage{caption} 
\usepackage{listings}

\definecolor{myblue}{rgb}{0.2,0.2,0.7}
 \usepackage{float}
\usepackage[
      colorlinks=true,
      linkcolor=myblue,
      urlcolor=blue,
      filecolor=blue,
      citecolor=red,
      pdfstartview=FitV,
      ]{hyperref}
\usepackage{subcaption}
\usepackage{cite}

\renewcommand{\thefootnote}{\fnsymbol{footnote}}

\definecolor{usuzumiiro}{cmyk}{0.08,0.05,0.06,0.55}

\usepackage{wrapfig}

\definecolor{tsuyukusairo}{cmyk}{0.9,0.36,0,0}
\definecolor{mizuasagi}{cmyk}{0.45,0.04,0.14,0}
\definecolor{moegi}{cmyk}{0.56,0.08,0.95,0}
\definecolor{hiwairo}{cmyk}{0.28,0.1,0.95,0}
\definecolor{wakabairo}{cmyk}{0.35,0,0.47,0.04}
\definecolor{kihadairo}{cmyk}{0.15,0.2,0.92,0}
\definecolor{kariyasuiro}{cmyk}{0.18,0.1,0.95,0}
\definecolor{uguisucha}{cmyk}{0.37,0.42,0.82,0.65}
\definecolor{araishu}{cmyk}{0,0.7,0.73,0.1}
\definecolor{botan'iro}{cmyk}{0.18,0.82,0,0}
\definecolor{usuzumiiro}{cmyk}{0.08,0.05,0.06,0.55}

\numberwithin{equation}{section}
\begin{document}

%
\begin{titlepage}
\begin{flushright}
\texttt{KIAS-P15010}\\
\texttt{RIKEN-STAMP-3}\\
\end{flushright}
\vspace{0.5cm}
\begin{center}
{\Large \bf
Tao Probing the End of the World
}
\\
\vspace{0.9cm}
$\,$Sung-Soo Kim$^{a}$\footnote{Email: sungsoo.kim@kias.re.kr},
$\,$ Masato Taki$^{b}$\footnote{Email: taki@riken.jp},
$\,$ and Futoshi Yagi$^{a}$\footnote{Email: fyagi@kias.re.kr}
\\
\vspace{0.3cm}
\it $^a$ Korea Institute for Advanced Study (KIAS)\\
85 Hoegiro Dongdaemun-gu, 130-722, Seoul, Korea
\\
\it $^b$  iTHES Research Group, RIKEN, Saitama 351-0198, Japan
\\
\vspace{0.3cm}
\end{center}
\vskip1.0cm

\begin{center}
\end{center}
\begin{abstract}
We introduce a new type IIB 5-brane description for the E-string theory which is the world-volume theory on the M5-brane probing the end of the world M9-brane.
The E-string in the new realization is depicted as spiral 5-branes web equipped with the cyclic structure which is key to uplifting to six dimensions. 
Utilizing the topological vertex to the 5-brane web configuration enables us to write down a combinatorial formula for the generating function of the E-string elliptic genera,
namely the full partition function of topological strings on the 
local $\frac{1}{2}$K3 surface.
\end{abstract}
\end{titlepage}


\renewcommand{\thefootnote}{\arabic{footnote}} \setcounter{footnote}{0}


\section{Introduction}

Among possible interacting superconformal field theories (SCFTs) in diverse dimensions,
the six-dimensional (6d) theories are less well understood.  
It is expected that 6d $\mathcal{N}=(1,0)$ SCFTs involve tensionless strings
and show very non-trivial physics. 
This non-trivial nature of the 6d SCFTs is deeply related to the strong coupling physics of superstring theory, but its detailed features have yet to be discovered. 

The E-string theory is a typical 6d $\mathcal{N}=(1,0)$ SCFT.
Originally, the E-string theory was found in the heterotic strings for the small $E_8$ instanton \cite{Witten:1995gx,Ganor:1996mu,Seiberg:1996vs}.
There are also several other realizations of the theory.
For instance, by considering M5-brane probing the end of the world M9 boundary \cite{Horava:1995qa,Horava:1996ma},
we can find the E-string theory on the M5 world-volume. 
M-theory or a topological string, on the local $\frac{1}{2}$K3 surface,
which describes a Calabi--Yau 3-fold in the vicinity of the $\frac{1}{2}$K3 4-cycle, is dual to the E-string theory  \cite{Klemm:1996hh}.
Shrinking a $\frac{1}{2}$K3 4-cycle in Calabi--Yau compactification of F-theory 
also leads to the E-string theory \cite{Witten:1996qb,Morrison:1996na,Morrison:1996pp}.
The Seiberg--Witten curve, the prepotential, and the partition function of the E-string theory
have been studied in \cite{Minahan:1997ct,Minahan:1998vr,Hosono:1999qc,Mohri:2001zz,Eguchi:2002fc,
Iqbal:2002rk,Eguchi:2002nx,Sakai:2011xg,Sakai:2012zq,Sakai:2012ik,Huang:2013yta,Ishii:2013nba,Sakai:2014hsa}.
Moreover, quite recently there was an attempt at classifying all
6d $\mathcal{N}=(1,0)$ SCFTs \cite{Heckman:2013pva,Gaiotto:2014lca,DelZotto:2014hpa,Heckman:2014qba,Haghighat:2014vxa}.
The anomaly polynomials of these 6d theories were also determined in \cite{Ohmori:2014pca,Ohmori:2014kda,Intriligator:2014eaa}.

Recently there has been significant progress on the localization computation in 5d gauge theories.
In \cite{Kim:2012gu,Hwang:2014uwa}, the localization calculation was formulated
for the superconformal index of 5d $\mathcal{N}=1$ $Sp(N)$ gauge theories with $N_f$ flavors,
and it was found that this index shows the enhancement of global symmetry
$SO(2N_f)\times U(1)\subset E_{N_f+1}$
as was expected from Seiberg's argument \cite{Seiberg:1996bd}. 
The same index was also derived from the topological string theory in \cite{Iqbal:2012xm},
and the enhancement was studied in more detail in \cite{Bao:2013pwa,Hayashi:2013qwa} 
from the perspective of topological strings on local Calabi--Yau 3-folds.
In \cite{Mitev:2014jza},
it was found that Nekrasov partition functions also show the enhancement.
These developments led to exact computation on the partition function and superconformal index of a 5d theory starting directly from the IR Lagrangian or from the corresponding brane web configuration/Calabi--Yau geometry.
This technology is the main ingredient of our approach to the E-string theory.

Thanks to the developments in localization,
quantitative study on 6d $\mathcal{N}=(1,0)$ SCFTs
is getting more manageable.
In \cite{Haghighat:2014pva,Kim:2014dza},
based on the elliptic genus computation, the E-string partition function up to four E-strings was computed. See also \cite{Cai:2014vka}.

In this paper we propose a different description of the E-string theory
based on the familiar type IIB 5-brane setup.
In the case of 5d $\mathcal{N}=1$ gauge theories, 
the world-volume theory of the corresponding $(p,q)$ 5-brane web configuration
leads to the desired gauge theory \cite{Aharony:1997ju,Aharony:1997bh,Leung:1997tw,Benini:2009gi}.
The same web diagram specifies the  toric Calabi--Yau 3-fold of the compactified M-theory dual to the 5-brane web,
and then we calculate the 5d Nekrasov partition function \cite{Nekrasov:2002qd} 
using the topological string method \cite{Aganagic:2002qg,Iqbal:2002we,Iqbal:2003ix,Iqbal:2003zz,Eguchi:2003sj,Taki:2007dh} 
known as the topological vertex \cite{Aganagic:2003db,Awata:2005fa,Iqbal:2007ii,Awata:2008ed,Iqbal:2012mt} through
geometric engineering \cite{Katz:1996fh,Katz:1997eq}.
If these techniques are also applicable to the 6d E-string theory,
they will be a very powerful method conceptually and computationally to investigate 6d dynamics.
To this end, we utilize the key fact: the E-string theory of interest, which is a 6d $\mathcal{N}=(1,0)$ SCFT, 
appears as the UV fixed point of {\it {5d}} $\mathcal{N}=1$ $SU(2)$ gauge theory with eight flavors.
Indeed, the agreement of the corresponding partition functions is checked in \cite{Kim:2014dza}. This means that once one finds a 5-brane web for $SU(2)$ gauge theory with eight flavors, one can apply all the established techniques to study the E-strings. 
In this paper, we discuss such $(p,q)$ 5-brane web description of the E-string theory. The $(p,q)$ 5-brane web that we found is of a spiral shape with a cyclic structure associated with the spiral direction, which is the source of the hidden 6d direction accounting for the Kaluza--Klein (KK) direction.
By implementing the topological vertex to this spiral web,
we can write down a combinatorial expression of the full-order partition function,
which is the generating function of the elliptic genera of the E-strings. 
This generating function is precisely the full partition function of topological strings on the local $\frac{1}{2}$K3 surface.
Our analysis may provide a stepping stone allowing one to look for a similar description for other 6d theories as well as a new class of 5-brane web.

The paper is organized as follows. In Sect. \ref{sec:tao 5 and 7branes}, we review type IIB $(p,q)$ 5-brane web construction for $SU(2)$ gauge theory with eight flavors based on 7-brane monodromies, and find a spiral structure of the web diagram, which we call the ``Tao diagram.'' In Sects. \ref{sec:Taopara} and \ref{sec:E-stringviaTopver}, applying the topological vertex method to this Tao diagram, we compute the partition function and compare the obtained result to the elliptic genera computed in \cite{Kim:2014dza}.  
We discuss various Tao diagrams as well as other future work in Sect. \ref{sec:discussion}.


\section{Tao 5-brane web via 7-branes}\label{sec:tao 5 and 7branes}

\subsection{7-brane background and 5d $\boldsymbol{SU(2)}$ gauge theories}
A large number of 5d $\mathcal{N}=1$ gauge theories and their UV fixed point SCFTs
are realized as the world-volume theories on 5-brane $(p,q)$ webs
\cite{Aharony:1997ju,Aharony:1997bh,Benini:2009gi,Bao:2011rc,Bao:2013pwa,Hayashi:2013qwa,Taki:2013vka,Bergman:2013aca,Taki:2014pba,Hayashi:2014wfa,Bergman:2014kza,Hayashi:2014hfa}.
This setup is precisely the 5d version of the Hanany--Witten brane configuration.
We can utilize this web realization to know the strongly coupled UV fixed point, 
the BPS spectra, the Seiberg--Witten curves and so forth.

In order to obtain the 5-brane $(p,q)$ webs for a given theory, one often finds it convenient to start from the system of 7-branes studied in \cite{DeWolfe:1999hj}. 
Technology for treating 7-brane backgrounds was developed in \cite{Gaberdiel:1997ud,Gaberdiel:1998mv,DeWolfe:1998zf,DeWolfe:1998eu,DeWolfe:1998pr}.
\begin{figure}[t]
 \begin{center}
\includegraphics[width=15cm, bb=0 0 971 310]{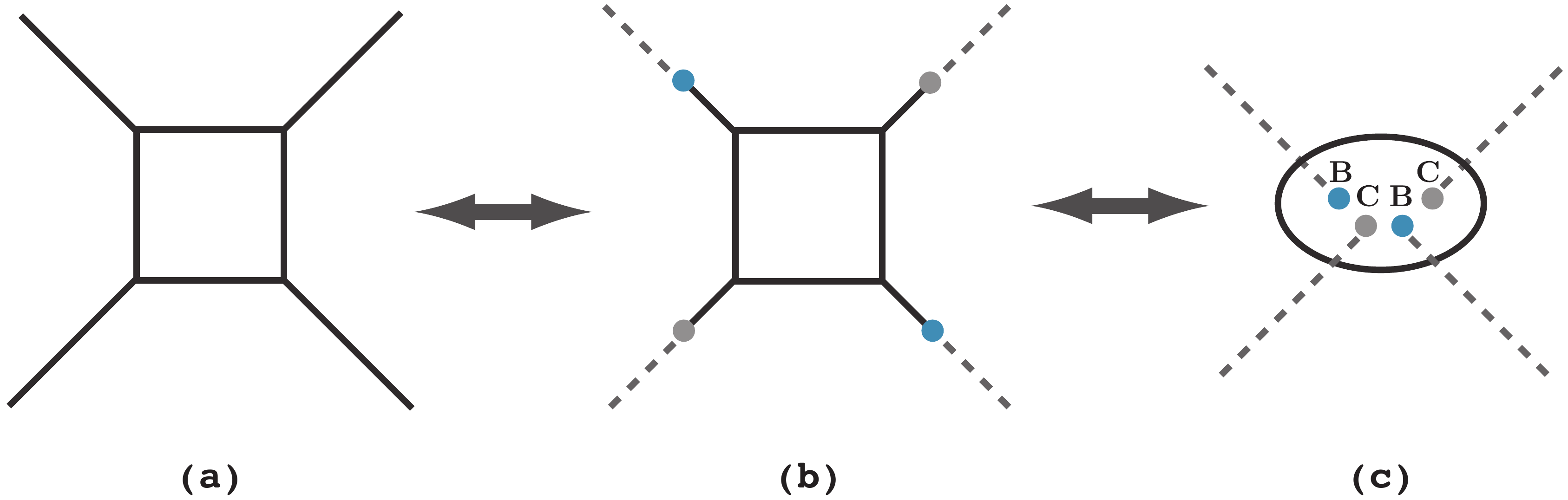}
 \end{center}
 \caption{\small  (a) A 5-brane web for $SU(2)$ pure Yang--Mills theory
 that corresponds to the local first del Pezzo $\mathbb{CP}^1\times \mathbb{CP}^1$.
 Introducing 7-branes (b) regularizes the configuration. The Hanany--Witten effect leads to (c) where the 7-brane background ${\bf B}{\bf C}{\bf B}{\bf C}$
 is probed by a 5-brane loop.}
 \label{fig:YM7brane}
\end{figure}
The relation between 7-brane background and 5-brane configuration is as follows. 
Figure \ref{fig:YM7brane} illustrates a typical example of the correspondence between a 5-brane web system 
and a 7-brane configuration.
Figure \ref{fig:YM7brane}(a) is the $(p,q)$ 5-brane web for 5d $\mathcal{N}=1$ $SU(2)$ pure Yang--Mills theory.
There are four external legs in this configuration, whose $(p,q)$ charges are $(1,1)$ and $(1,-1)$.
We can regularize these semi-infinite external legs by terminating a $(p,q)$ leg on a $[p,q]$ 7-brane,
and then we obtain Figure \ref{fig:YM7brane}(b).
Without changing the world-volume theory, these 7-branes illustrated by colored circles can move freely along $(p,q)$ line.
We can therefore move them inside the 5-brane quadrilateral, 
and then the configuration in Figure \ref{fig:YM7brane}(c) appears.
Notice that the four external legs have disappeared 
since the Hanany--Witten brane annihilation transition removes these 5-branes 
after the 7-branes cross the 5-brane quadrilateral.
The 5-brane quadrilateral are now highly curved because of the non-trivial metric coming from the 7-brane inside.

\begin{figure}[t]
 \begin{center}
\includegraphics[width=16cm, bb=0 0 2204 602]{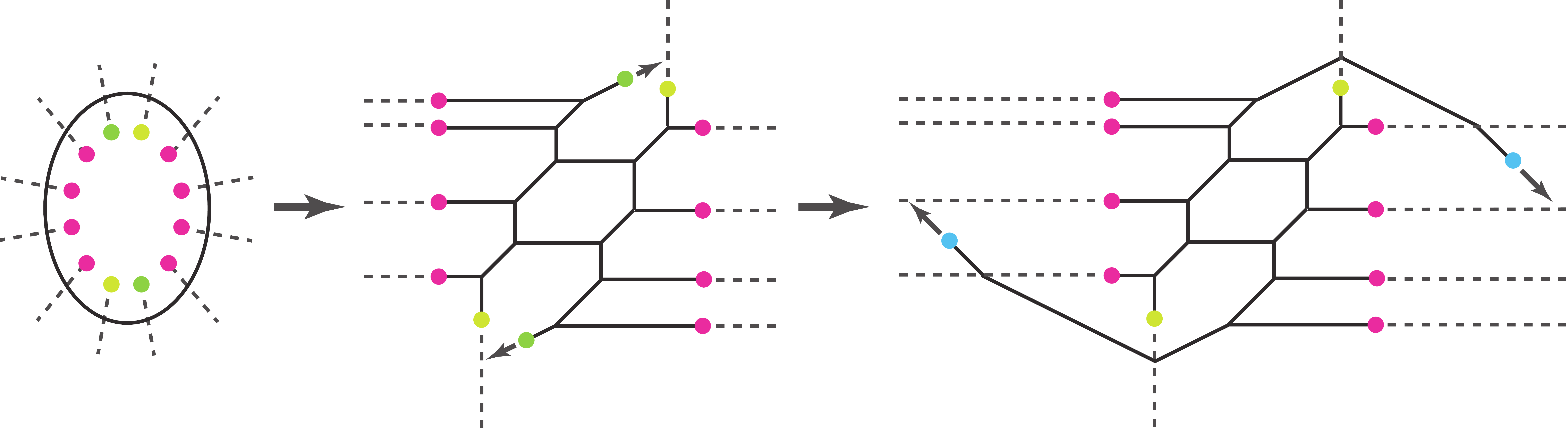}
 \end{center}
 \caption{\small  On the left-hand side, the starting configuration with 12 7-branes in the 5-brane loop.
Moving them outside leads to the middle diagram (up to flop transition). On the right-hand side, 
 we can move two 7-branes (blue dots) to infinity, which makes a diagram with two arms that rotate in a spiral infinitely many times.}
 \label{fig:HowToMakeTao}
\end{figure}

In summary, the $(p,q)$ brane web for $SU(2)$ Yang--Mills theory
is related to a 5-brane loop configuration probing the 7-brane configuration 
\begin{align}
\hat{\bf E}_1
={\bf B}{\bf C}{\bf B}{\bf C},
\end{align}
where ${\bf B}$ is a $[1,-1]$ 7-brane and ${\bf C}$ is a $[1,1]$ 7-brane.
The world-volume theory on this 5-brane loop is also $SU(2)$ Yang--Mills theory
because of the above construction.
In this way, we can convert a $(p,q)$ web into a 5-brane loop probing a 7-brane background.
For 5d $\mathcal{N}=1$ $SU(2)$ gauge theory with $N_f$ flavors,
the following 7-brane background appears inside the 5-brane loop:
\begin{align}
\label{affine7branebkgd}
\hat{\bf E}_{N_f+1}
={\bf A}^{N_f}{\bf B}{\bf C}{\bf B}{\bf C}.
\end{align}
Here, ${\bf A}$ is a $[0,1]$ 7-brane.
Up to seven flavors, the 7-brane system can be recast into a 5-brane web configuration 
by pulling out all the 7-branes to infinity using Hanany--Witten transition
\cite{Aharony:1997bh,DeWolfe:1999hj,Benini:2009gi,Taki:2014pba}. 
In the language of the toric Calabi--Yau
associated with the corresponding $(p,q)$ web diagram \cite{Leung:1997tw},
this 7-brane configuration is dual to the local $\mathbb{P}^2$  blown up at $N_f+1$ points or local $\mathbb{P}^1\times\mathbb{P}^1$ 
geometry, namely the local del Pezzo surfaces $\mathcal{B}_{N_f+1}$.

Up to the $N_f=4$ flavors, it is easy to see the corresponding $(p,q)$ 5-brane web diagrams (See the appendix of \cite{Kim:2014nqa}). It is, however, not so straightforward to find $(p,q)$ 5-brane web diagrams for $N_f=5,6,7$ flavors. It was studied in 
\cite{Benini:2009gi,Bao:2013pwa,Hayashi:2013qwa} that the corresponding 7-brane configurations for $N_f=5,6,7$ flavors are realized in the $(p,q)$ 5-brane web diagrams as 5d uplifts of tuned (Higgsed) $T_N$ diagrams : The 7-brane configuration for the $N_f=5$ flavor case is exactly mapped to the $T_3$ diagram showing $E_6$ symmetry; the configurations for the $N_f=6,7$ flavor cases are tuned $T_4$ and $T_6$ diagrams showing $E_7$ and $E_8$ symmetry, respectively.

\begin{figure}
\centering
\begin{subfigure}{.5\textwidth}
  \centering
  \includegraphics[width=7.5cm, bb=0 0 1484 1075]{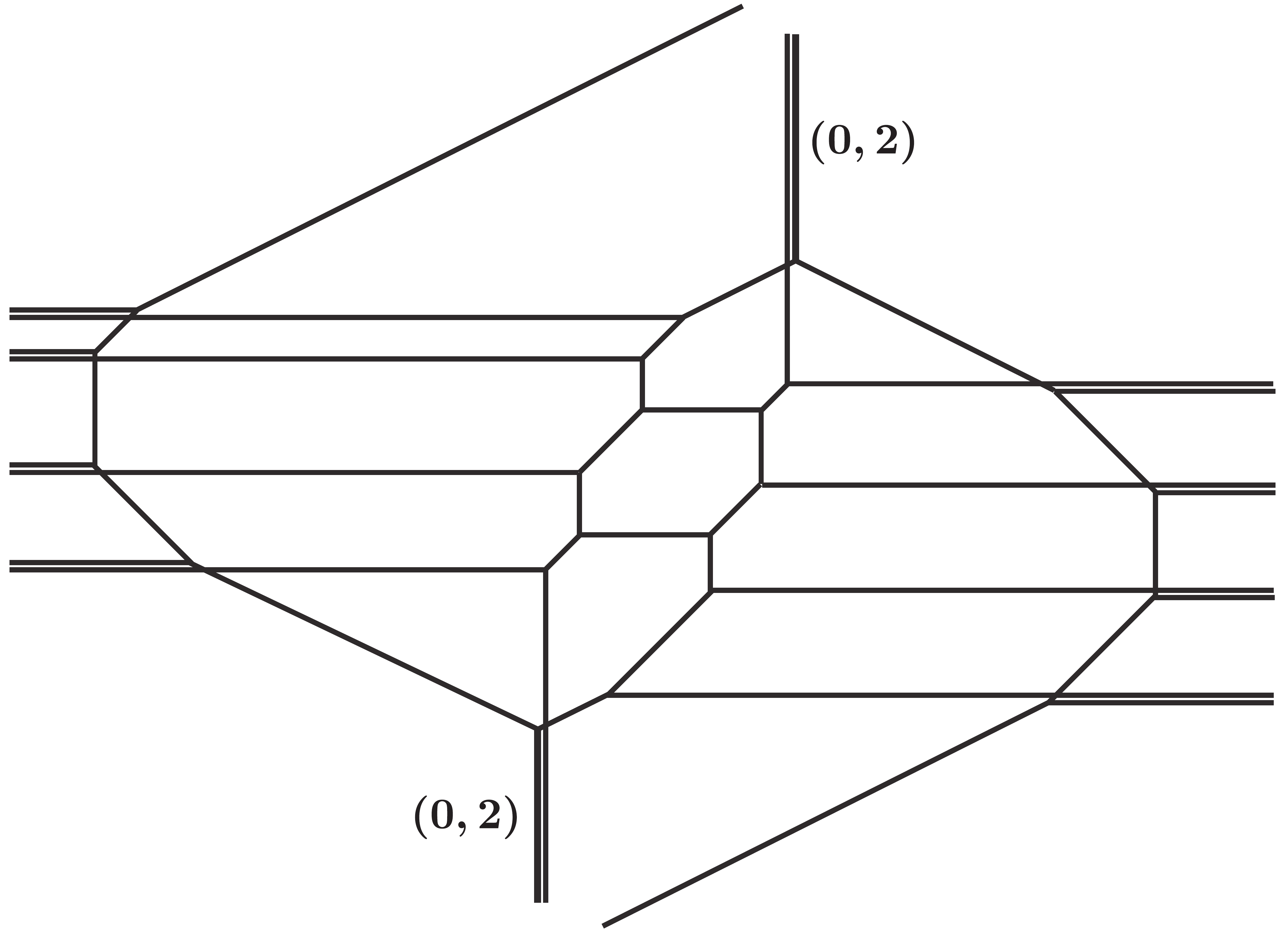}
   \caption{\small  A trial Tao web associated with \\$SU(2)$ gauge theory with $N_f=8$ flavors.}
  \label{fig:Nf8}
\end{subfigure}%
~~~
\begin{subfigure}{.5\textwidth}
  \centering
  \includegraphics[width=25mm, bb=0 0 507 1075]{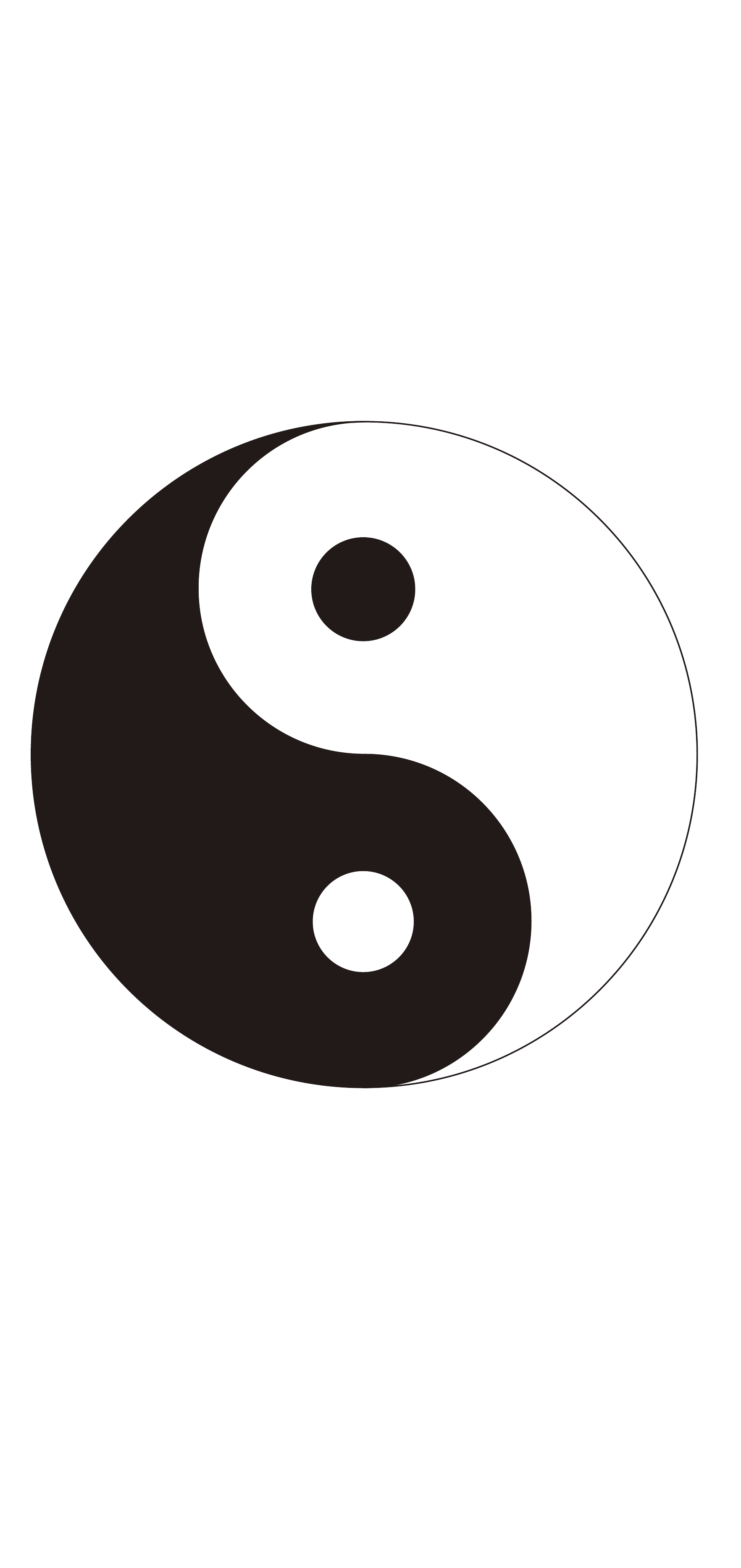}
   \caption{\small Yin-yang that is a symbolic representation of Taoism.}
 \label{fig:YY}
\end{subfigure}
\caption{\small  Similarity between a Tao web diagram and a Taoism symbol}
\label{fig:test}
\end{figure}

Our focus in this paper is the 5d $\mathcal{N}=1$ $SU(2)$ gauge theory with $N_f=8$ flavors
and its 6d UV fixed point theory, namely the the E-string theory.
Our starting point is therefore the affine 7-brane background (\ref{affine7branebkgd}) for $N_f=8$.
This 7-brane configuration corresponds to the $\frac{1}{2}$K3 surface,
and M-theory or topological strings on the Calabi--Yau is dual to the E-string theory.
The $N_f=8$ configuration is a one-point blow-up of the $N_f=7$ configuration. 
Using 7-brane monodromies one easily finds that 
\begin{align}
\label{nf9config}
\hat{\bf E}_{9}
={\bf A}^{8}{\bf B}{\bf C}{\bf B}{\bf C}={\bf A}^4 {\bf X}_{[2,1]} {\bf N} {\bf A}^4 {\bf X}_{[2,1]} {\bf N},
\end{align}
where $ {\bf N}$ is a $[0,1]$ 7-brane and ${\bf X}_{[p,q]}$ is a $[p,q]$ 7-brane.
See Appendix \ref{app:Convention} for more detail.
The corresponding 5-brane loop on this 7-brane background is illustrated in Figure \ref{fig:HowToMakeTao}(a).
 \begin{figure}[t]
 \begin{center}
\includegraphics[width=8cm, bb=0 0 494 191]{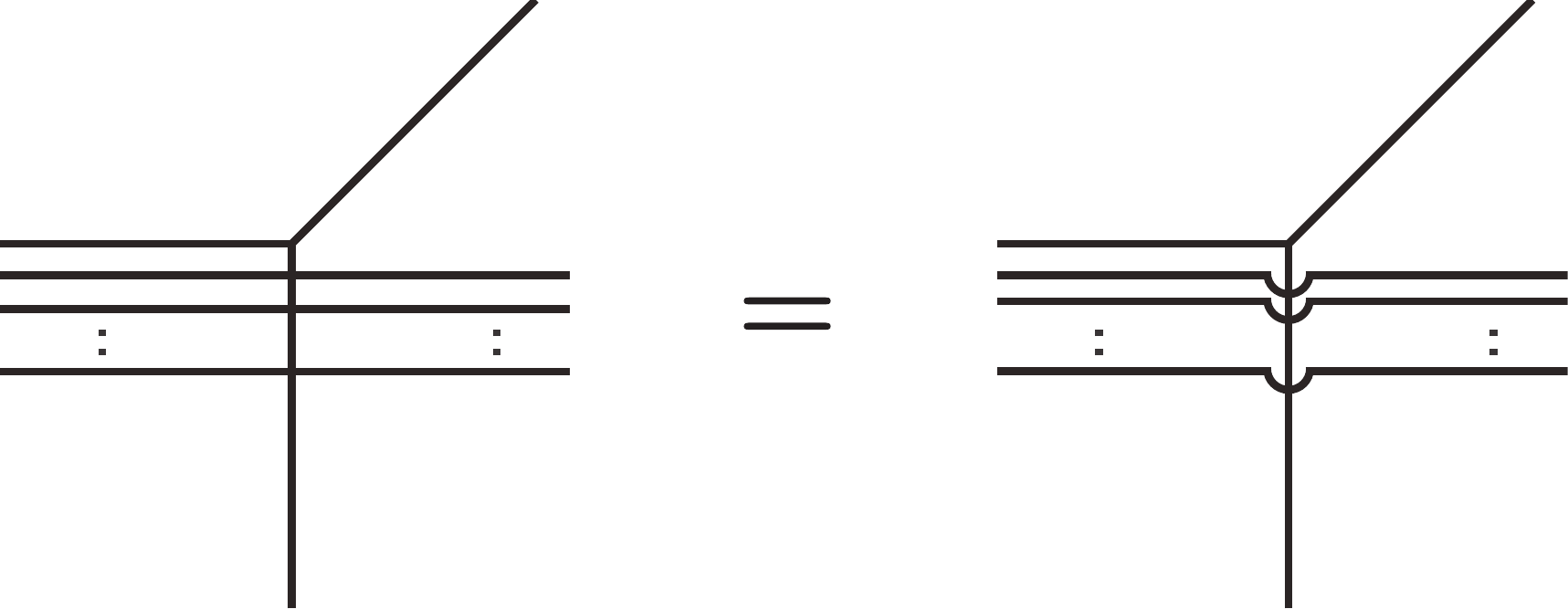}
 \end{center}
 \caption{\small Straight 5-branes crossing a 5-brane are introduced to denote the 5-branes that are actually jumping over a 5-brane, for simplicity.}
 \label{fig;jump}
\end{figure}
When we try to pull out the 7-branes, we find that the $[2,1]$ 7-branes change their charges
due to the monodromy cut created by $[1,0]$ 7-branes and $[0,1]$ 7-branes, 
which makes the attached 5-branes form spiral arms.
 After taking all 7-branes to infinity, we find a spiral  
 web diagram with multiple arms which rotate infinitely many times, as in Figure 
 \ref{fig:test}(a).
We will call such a spiral diagram ``Tao diagram'' for short
because its spiral structure is close to a symbolic representation 
of the Taoism philosophy Figure \ref{fig:test}(b).
This Tao diagram is an intuitive one that reveals the spiral nature of the $(p,q)$ 5-brane web diagram for $N_f=8$ flavors. 

Note that Figure \ref{fig:test}(a) contains multiple coincident 5-branes intersecting at one vertex.
These 5-branes actually jump over the other 5-brane at such a vertex in the sense of \cite{Benini:2009gi}. 
For instance, see Figure \ref{fig;jump}.
In topological vertex formalism, such jumping is realized by a degenerate K\"ahler parameter 
\cite{Kozcaz:2010af,Dimofte:2010tz,Taki:2010bj,Aganagic:2012ne,Hayashi:2013qwa}.
We will show definite way to treat it in the next section.

Unfortunately, the above Tao web Figure \ref{fig:test}(a) involves strange ``$(0,2)$ 5-branes'' which can
not be any bound state.
To avoid discussion of the proper treatment of such 5-branes,
we switch to a more healthy web description in the following. 
In fact, a web description for a given 7-brane background
is not unique because changing the ordering of 7-brane movement results in a different web \cite{Taki:2014pba}.
In the next subsection, we demonstrate in detail that there exists another Tao web for the E-string theory,
and we will use this new Tao diagram throughout this paper.

\subsection{E-string theory via Tao $\boldsymbol{(p,q)}$ web}
As we have observed in the previous subsection,
$N_f=8$ theory leads to spiral 5-brane configuration
and opens up a new dimension associated with the cyclic nature. 
Such a Tao web would give an intuitive and effective description of the 6d E-string theory. 
In this subsection, we give a simpler and more useful Tao diagram which describes the E-string theory.
The starting point is the affine 7-brane background again,
\begin{align}
\hat{\bf E}_9
={\bf A}^8\,{\bf B}{\bf C}{\bf B}{\bf C}.
\end{align}
We will consider the 5-brane loop configuration on this configuration to construct a dual description
of the local $\frac{1}{2}$K3 surface
as was done in the cases of the local del Pezzo surface \cite{DeWolfe:1999hj,Benini:2009gi,Bao:2013pwa,Hayashi:2013qwa,Taki:2014pba}.
The resulting Tao web is therefore a generalized toric description of the local $\frac{1}{2}$K3 surface.

\begin{figure}[t]
 \begin{center}
\includegraphics[width=13cm, bb=0 0 1352 612]{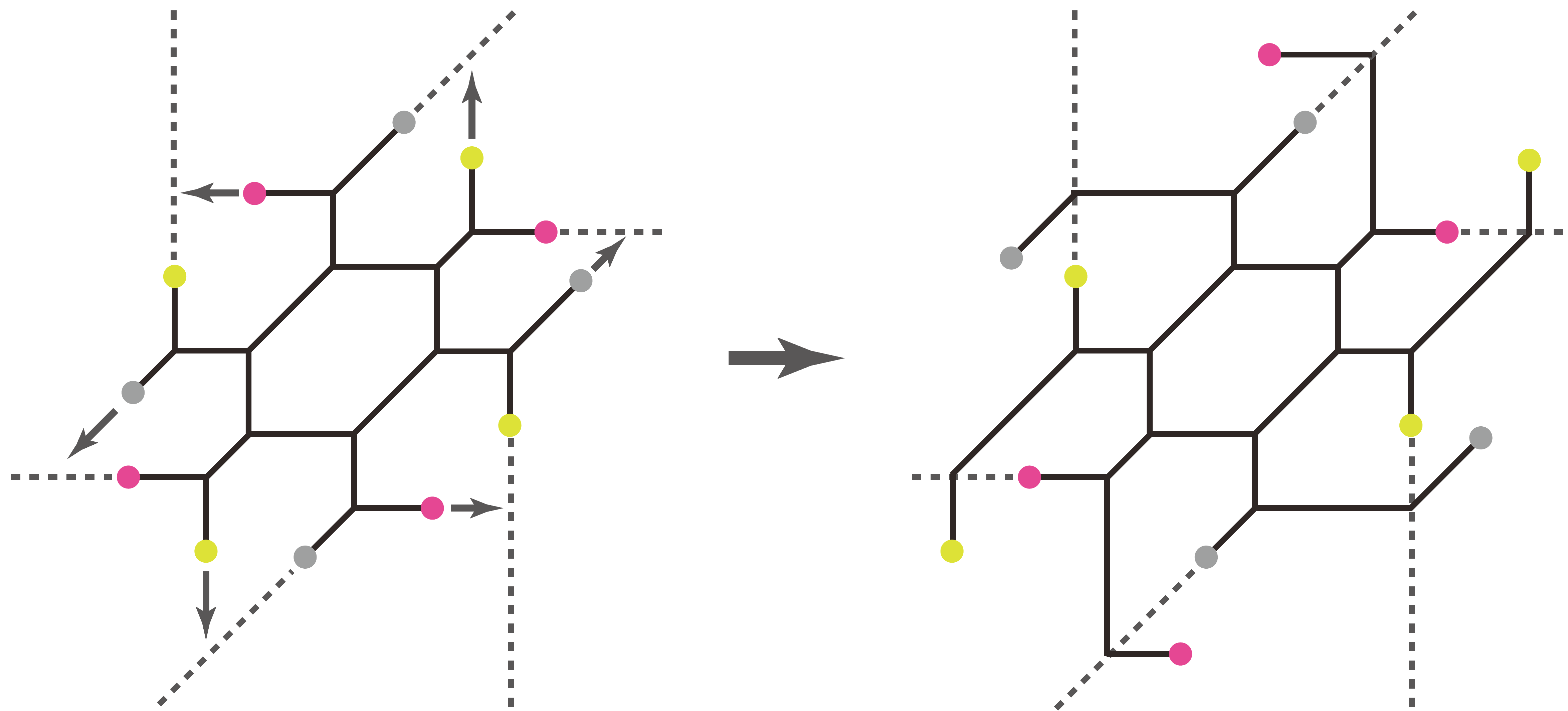}
 \end{center}
 \caption{\small The 7-brane motion that leads to the 5-brane Tao web.}
 \label{fig:7branemove}
\end{figure}

\begin{figure}[t]
 \begin{center}
\includegraphics[width=15cm, bb=0 0 2221 1028]{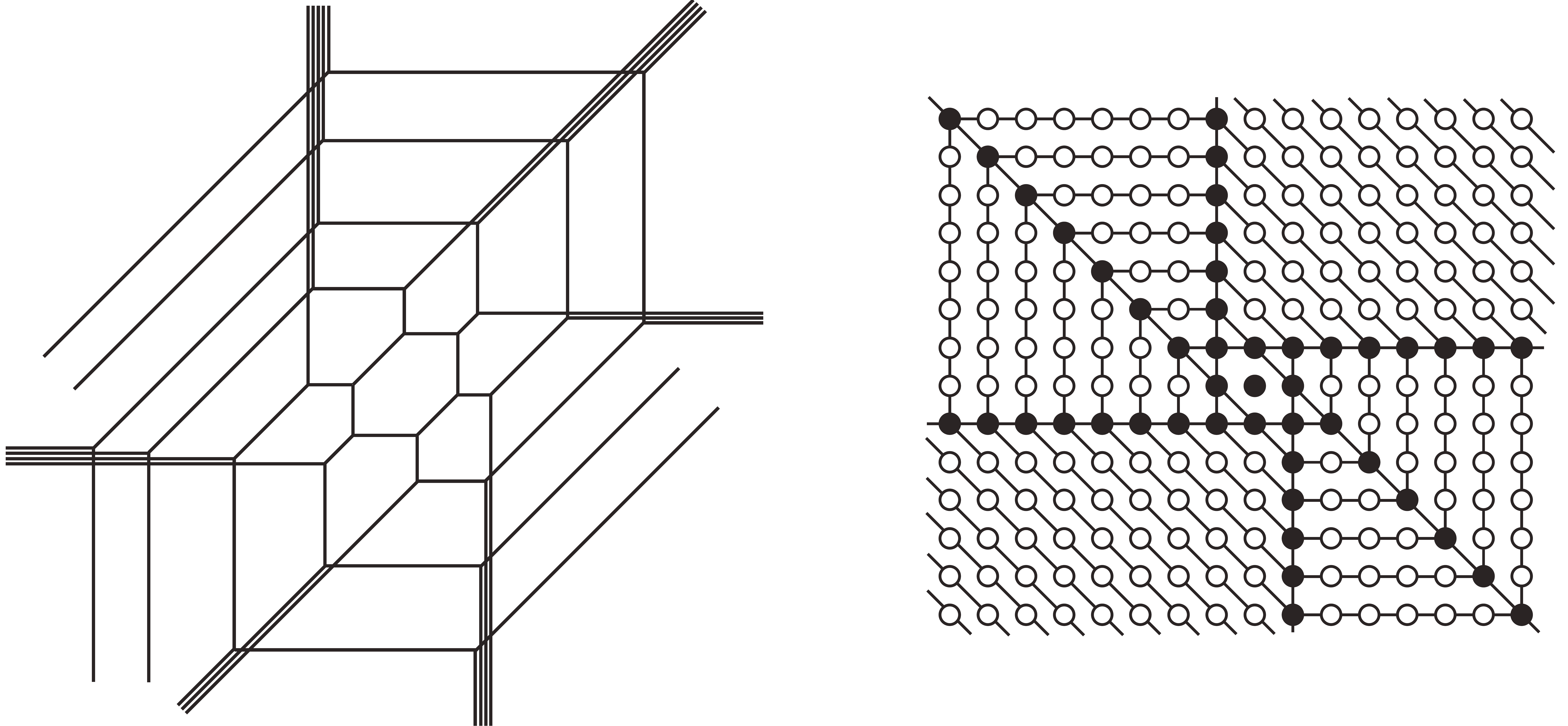}
 \end{center}
 \caption{\small A 5-brane Tao web and its black and white dots diagram.
 This is a ``toric-like'' description of the E-string theory.} 
 \label{fig:taoweb}
\end{figure}

As is explained in Appendix \ref{app:Convention}, reordering these 7-branes leads to another expression of the $\hat{\bf E}_9$ configuration:
\begin{align}
\label{Tao7branes}
\hat{\bf E}_9
={\bf A}{\bf N}{\bf C}\,{\bf A}
{\bf N}{\bf C}\,{\bf A}{\bf N}
{\bf C}\,{\bf A}{\bf N}{\bf C}.
\end{align}
\begin{figure}[t]
 \begin{center}
\includegraphics[width=14cm, bb=0 0 1447 427]{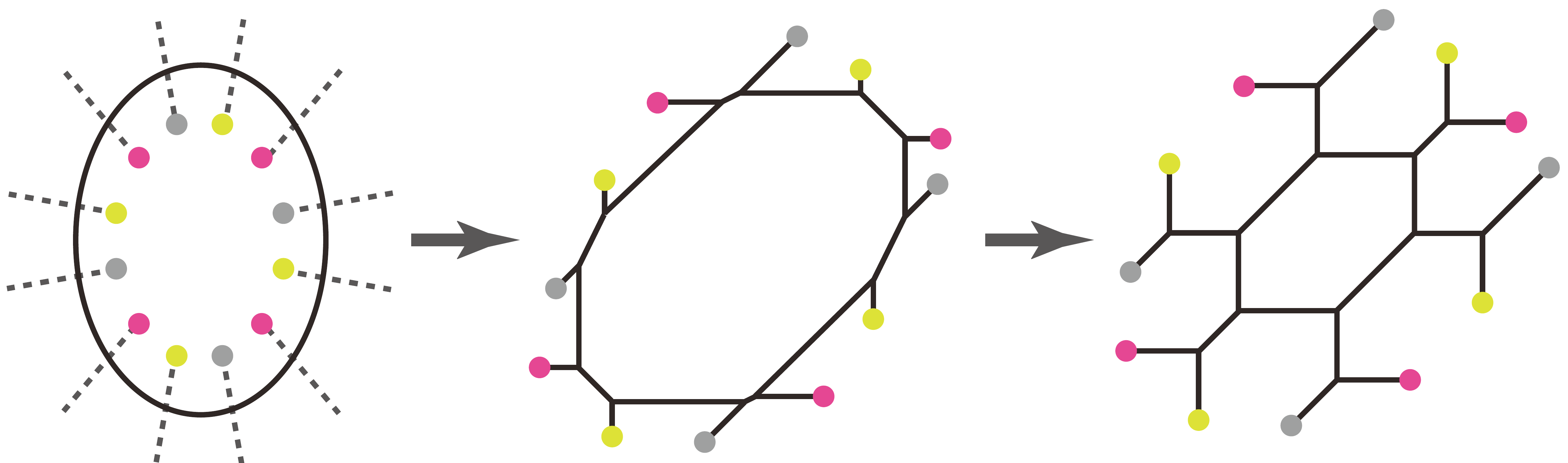}
 \end{center}
 \caption{\small A 5-brane web configuration that gives the local ninth del Pezzo surface.}
 \label{fig:7braneE9}
\end{figure}
This configuration probed by a 5-brane loop is the left-hand side of Figure \ref{fig:7braneE9}. The solid
black line is the 5-brane loop probing 7-branes,
and all the branch-cuts illustrated with the gray dashed lines are outgoing.
Pulling out the 7-branes from the 5-brane loop together with the flop transitions
leads 
to  the right-hand side of Figure \ref{fig:7braneE9}.
The Hanany--Witten effect introduces new 5-brane prongs attached to the 7-branes.
We can  continue to move the 7-branes away from the middle 5-brane loop;
however, a 7-brane soon hits a branch-cut coming from another 7-brane.
The shape of the resulting web configuration  therefore depends on the ordering of the 7-brane motion.
Let us consider the 7-brane motion illustrated in
Figure \ref{fig:7branemove}.
Continuing to move 7-branes with such an ordering,
we finally find the spirally growing 5-brane web in Figure \ref{fig:taoweb}.
This is precisely the Tao brane web which describes $SU(2)$ gauge theory with $N_f=8$ flavors,
that is, the E-string theory.

This Tao web in Figure \ref{fig:taoweb} has six external legs making spiral cycles,
and each leg has period 6 of its joints.
In addition, there are six straightly outgoing legs which consist of bundles of 5-branes of the same $(p,q)$ type.
A bundle increases the number of constituent 5-branes by one each time the bundle crosses a spiral leg.
This structure comes from the Hanany--Witten effect that occurs when one moves all remnants of 7-branes to infinity
after creating the six spiral legs in Figure \ref{fig:7branemove}.
This local structure is a degenerate version of the $T_{N\to\infty}$ 5-brane web.
From this point of view,
we can construct the Tao web by gluing together six $T_{\infty}$ sub-webs. 
Treatment of such a degenerate toric diagram was studied in \cite{Kozcaz:2010af,Dimofte:2010tz,Taki:2010bj,Aganagic:2012ne,Hayashi:2013qwa}.
We will explain it in the next section.

\subsection{What will happen when $\boldsymbol{N_f\geq9}$?}

It is claimed in \cite{Seiberg:1996bd} that 
5d $Sp(1)$ gauge theory with $N_f$ flavors has 5d UV fixed point only for $N_f \le 7$
and that it does not for $N_f\geq 9$.
The E-string case $N_f=8$ is critical, where it has 6d UV fixed point.
However, it looks as if we can consider $N_f\geq 9$ configuration at least formally starting from the 7-brane background (\ref{affine7branebkgd}).
Does our brane web system ``know'' that $N_f=8$ is critical?
The $N_f\leq 7$ cases lead to finite web diagrams associated with the local del Pezzo surfaces; the $N_f=8$ configuration opens up a new $S^1$ direction, and it gives the Tao diagram associated with the local $\frac{1}{2}$K3 surface.
Then, what will happen for $N_f\geq 9$?

A way to see that the $N_f=8$ case is critical is that if one successively applies 7-brane monodromy to move a $[p,q]$ 7-brane through other 7-branes, then it comes back to the original configuration after one rotation, which is a necessary condition for the spiral shape of the 5-brane web. For instance, the 7-brane configuration corresponding to the middle diagram in Figure \ref{fig:HowToMakeTao} is given by
\begin{align}
\label{eq:7braneconf4fig3}
{\bf X}_{[2,1]}{\bf N}{\bf A}^4{\bf X}_{[2,1]}{\bf N}{\bf A}^4,
\end{align}
and two $[2,1]$ 7-branes (green dots in the diagram) are rotating clockwise. This means that these two 7-branes undergo the following monodromies due to the branch cuts from the remaining 7-branes:
\begin{align}
{\bf N}{\bf A}^4{\bf N}{\bf A}^4,
\end{align}
hence the charges of the rotating 7-branes changes as it passes through each cut. For one rotation,
the monodromy matrix that two $[2,1]$ 7-branes go through is  
\begin{align}
K=K_{[1,0]}^{-4}
K_{[0,1]}^{-1}
K_{[1,0]}^{-4}
K_{[0,1]}^{-1} = \begin{pmatrix}
5~&~-8\cr
2~&~-3
\end{pmatrix},
\end{align}
which yields 
\begin{align}
\begin{pmatrix}
5~&~-8\\
2~&~-3
\end{pmatrix}
\begin{pmatrix}
2\\
1
\end{pmatrix}
=\begin{pmatrix}
2\\
1
\end{pmatrix}.
\end{align}
Hence, the charge of two $[2,1]$ 7-branes remain the same after one rotation.

One can apply the same logic to $N_f=9$ flavors. A configuration for $N_f=9$ that one can find is one which adds one more flavor $[1,0]$ 7-branes to Figure \ref{fig:HowToMakeTao}:
\begin{align}
{\bf X}_{[2,1]}{\bf N}{\bf A}^5{\bf X}_{[3,1]}{\bf N}{\bf A}^4.
\end{align}
If we take the $[3,1]$ 7-brane and let it go clockwise through all the branch cuts then it is easy to see that the charge changes due to monodromy can never be same as any of the 7-branes above. In a similar way, the charge of a $[3,1]$ 7-brane becomes $[10,3]$ after one rotation, given by
\begin{align}
K_{[1,0]}^{-5}
K_{[0,1]}^{-1}
K_{[1,0]}^{-4}
K_{[0,1]}^{-1} \cdot [3,1] = [10,3].
\end{align}
Note that the direction that the resultant charge points is inward rather than outward. The more it goes around, the more it points inward. The 5-brane attached to this inwardly rotating 7-brane eventually crosses into the 5-brane loop in the middle, and thus it shrinks to the origin rather than going away from the origin. In other words the configuration with $N_f=9$ flavors never makes a proper 5-brane web, it all collapses. In a similar fashion, one finds collapsing of brane configuration for higher flavors $N_f>9$. This is consistent with the known result and it is a geometric account of why $N_f=8$ configuration is critical.

\section{Physical parameters of the Tao web}\label{sec:Taopara}

In this section, we go back to the Tao diagram, which corresponds to $N_f=8$. 
It is discussed in \cite{Leung:1997tw} that a $(p,q)$ 5-brane web can be reinterpreted as a toric diagram.
There are various reports \cite{Hayashi:2013qwa, Hayashi:2014wfa,Mitev:2014jza} 
that this claim also works for the ``toric-like diagram''\footnote{It was termed a ``dot diagram'' in their paper.}  introduced in \cite{Benini:2009gi}, which contains 5-branes jumping over other 5-branes.
Therefore, we expect that it is also the case for our Tao diagram even if it extends to infinity.
In other words, we expect that our Tao diagram gives a toric-like description for local $\frac{1}{2}$K3.
Assuming this, we compute 
the E-string partition function
using the topological vertex in the next section.
As a preparation, we need to study the relations among the K\"ahler parameters of the corresponding geometry
as well as their relation with the gauge theory parameters.

For each segment  $E\in\{\textrm{edges}\}$ in the web diagram, we associate the length parameter $t_E$
and exponentiated one $Q_E=\exp (-t_E)$.
In the language of toric-like geometry,
this parameter is the K\"ahler parameter of the two-cycle corresponding to the segment.
Unlike a usual web diagram,
there are infinitely many segments and their K\"ahler parameters in our diagram.
They are, however, highly constrained 
since the six radial legs trigger the turns of the spirals.
We can solve such constraint equations and finally find only ten free parameters.

We introduce a notation for organizing infinitely many K\"ahler parameters.
In the Tao web there are six spiral external legs,
and we label them by integers $\ell=1,2,\ldots,6$ in counterclockwise order.
An infinite number of straight-line segments compose a spiral leg.
To describe this spiral curve, we label K\"ahler parameters $Q_{i(\ell)}$, with $i$ being the turn number of the $\ell$th arm.
It is convenient to introduce the ``distance'' $\Delta_\ell$ between adjacent arms,
\begin{align}
Q_{i+1(\ell)} = \Delta_\ell\,Q_{i(\ell+1)},
\label{eq:DeltaQ}
\end{align}
where $\ell=1,2, \ldots, 6$, and $i=1,2,\ldots$.
We define
\begin{align}
\Delta_{\ell+6}=\Delta_{\ell},\qquad Q_{i(\ell+6)}=Q_{i(\ell)},
\end{align}
\begin{figure}[t]
 \begin{center}
\includegraphics[width=16cm, bb=0 0 3907 2023]{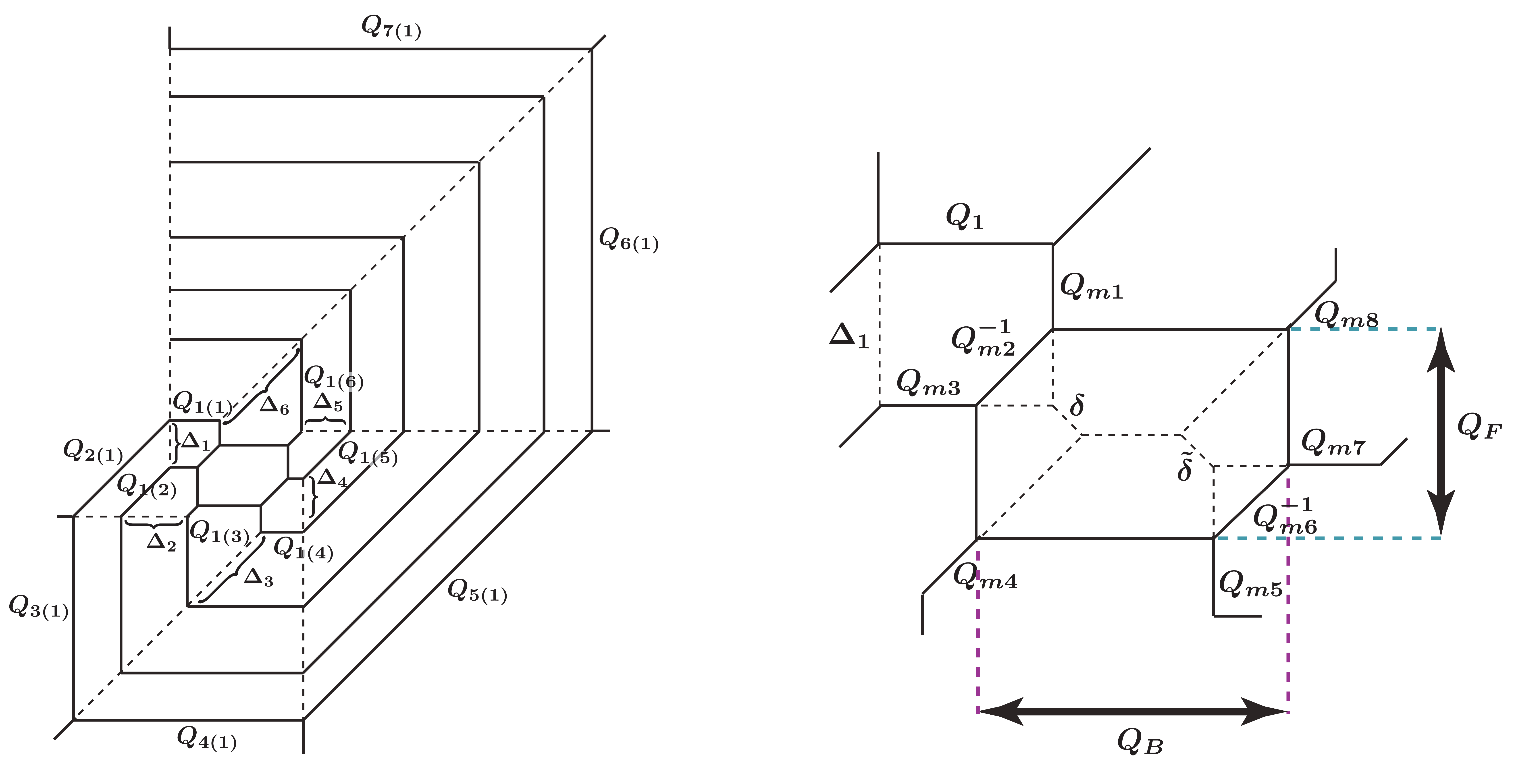}
 \end{center}
 \caption{\small The K\"ahler parameters of the Tao web.}
 \label{fig:notation}
\end{figure}
so that \eqref{eq:DeltaQ} is also satisfied for $\ell=0$ and/or $\ell=6$.
This parametrization is depicted in Figure \ref{fig:notation}.
By iteration, one finds that any K\"ahler parameter can be expressed as
\begin{align}
Q_{i(\ell)} = \bigg(\prod^{i-1}_{\ell'=1}\Delta_{\ell-1+\ell'}\bigg) Q_{1(\ell+i-1)}.
\end{align}
By introducing the ``period'' $d$ 
\begin{align}\label{period}
d=\prod^{6}_{\ell=1}\Delta_\ell,
\end{align}
one also finds that the $\ell$th arms are aligned parallel every six turns:
\begin{align}
Q_{i+6(\ell)}=d\cdot Q_{i(\ell)}.
\end{align}
We thus have 12 parameters: $\Delta_\ell$ and $Q_{1(\ell)}$. We note, however, that they depend only on nine out of ten physical parameters, eight flavors, and one gauge coupling.\footnote{In the Tao diagram, we have 
the K\"ahler moduli parameters $Q_F$, $Q_B$, and $Q_{mf}$, which parametrize the structure of the middle part of the diagram, as well as $Q_{i(\ell)}$, which are associated with the spiral legs, as depicted in Figure \ref{fig:notation}.
They depend on the ten physical parameters in total: eight mass parameters, one gauge coupling constant, and one Coulomb moduli parameter. However, it turns out that $Q_{i(\ell)}$ do not depend on the Coulomb moduli parameter and thus depend only on nine parameters. On the other hand, the remaining K\"ahler moduli parameters $Q_F$, $Q_B$, and $Q_{mf}$ do depend on the Coulomb moduli parameter as we will explain later.}
 Hence, the 12 parameters are not all independent, but are subject to the following three constraints:\\
(i) vertical constraint ($Q_{6(1)}Q_{5(1)} = d\, Q_{2(1)}Q_{3(1)}$):
\begin{align} 
\frac{Q_{1(2)}\,Q_{1(3)} }{Q_{1(5)}\,Q_{1(6)}} 
\, = \,\frac{\Delta_3\,\Delta_4}{\Delta_6\,\Delta_1} \ ,
\end{align}
(ii) horizontal constraint ($Q_{5(1)}Q_{4(1)} = d\, Q_{1(1)}Q_{2(1)}$):
\begin{align}
\frac{Q_{1(1)}\,Q_{1(2)} }{Q_{1(4)}\,Q_{1(5)}} 
\, = \,\frac{\Delta_2\,\Delta_3}{\Delta_5\,\Delta_6} \ ,
\end{align}
(iii) constraint on the origin of the Coulomb branch:
\begin{align}\label{eq:cons3}
\prod^6_{\ell=1} Q_{1(\ell)} = \prod^6_{\ell=1} \Delta_\ell. 
\end{align}
The last constraint requires some explanation. If we perform local deformation to make the vanishing Coulomb moduli (see the right-hand side of Figure \ref{fig:notation}), then the origin of the Coulomb branch is supposed to be the starting point of the K\"ahler  parameters $\delta$ and $\tilde\delta$. By taking the  horizontal projection associated with $\delta$, one finds that 
\begin{align}\label{eq:delta}
Q_{1(1)} \, Q_{1(2)}\,\delta = \Delta_2\, Q_{1(2)}\,\delta^{-1}\, \quad \Rightarrow \quad \delta^2 =\Delta_2\, Q_{1(1)}^{-1}. 
\end{align}
Likewise, one obtains
\begin{align}\label{eq:tdelta}
\tilde\delta^2 = \Delta_5\, Q_{1(4)}^{-1}. 
\end{align}
From the vertical projection, it is easy to see that
\begin{align}\label{eq:deltdel}
\delta\,\tilde\delta = Q_{1(2)}\,Q_{1(3)}\, \Delta_3^{-1}\,\Delta_4^{-1}. 
\end{align}

It follows from \eqref{eq:delta},\eqref{eq:tdelta}, and \eqref{eq:deltdel} 
that one finds \eqref{eq:cons3}. Therefore, the parameters $\Delta_\ell$ and $Q_{1(\ell)}$, constrained by (i), (ii), and (iii), can be a set of building blocks describing all the K\"ahler parameters for the Tao spiral.

Since only ten parameters are independent in the Tao web,
all the K\"ahler parameters can be written in terms of $Q_{mf}$, $Q_F$, and $Q_B$ in Figure \ref{fig:notation}. 
These bases are usually used for constructing a Nekrasov partition function in the context of a topological string.
For considering the flavor symmetry of the topological string partition function,
more useful parametrization is given by the $SO(16)$ fugacities corresponding to the masses $y_{f}=e^{-R m_f}$ ($f=1,2,\ldots8$),  
the instanton factor 
$\mathfrak{q}$ 
and the Coulomb moduli $A=e^{-Ra}$ 
introduced through the relations:
 \begin{align}
 Q_{mf}& = A^{-1}\,y_f  \quad (f=1,2,3,\ldots,8), \cr
 Q_{F} &= A^{2},  \label{eq: 3 relation}\\
 Q_{B} &= \frac{A^2 }{\prod^8_{f=1} y_f{}^{\frac12} } \mathfrak{q}. \nonumber 
\end{align}
The first and the second relations of \eqref{eq: 3 relation} are straightforward to understand if we carefully follow the sequence of the Hanany--Witten transition from Figure \ref{fig:HowToMakeTao} to Figure \ref{fig:7branemove}
and study which parameter in one diagram corresponds to which parameter in the other diagram.
We will see that $Q_{mf}$ corresponds to the 
distance between one of the color D5 branes and one of the flavor D5-branes,
while $Q_F$ corresponds to the distance between the two color D5-branes
in Figure \ref{fig:HowToMakeTao}.
We need further explanation for the third relation.
It is natural that the horizontal distance $Q_B$ is proportional to the instanton factor $\mathfrak{q}$
since the distance between two NS5-branes corresponds to the gauge coupling constant. 
On top of that, the prefactor in front of the right-hand side of $Q_B$ appears as follows:
When we compute the topological string partition function, the factor in the form $(1 - q^n Q)$ typically appears,
where $Q$ is a certain product of the K\"ahler parameters. 
When we rewrite this factor in the form of $\sinh$, we obtain the prefactor $\sqrt{q^n Q}$.
In order for the topological string amplitude to agree with the Nekrasov partition function,
we need to absorb the collection of such factors into $Q_B$ and regard it as the instanton factor $\mathfrak{q}$.
For $0 \le N_f \le 7$, it is explicitly checked that such a factor\footnote{For instance, see \cite{Bao:2011rc} for $N_f=4$ flavors.} 
is given by (the inverse of) $A^2 \prod^{N_f}_{f=1} y_f{}^{-\frac12}$.
We assumed that it is also the case for $N_f=8$.

From Figure \ref{fig:notation}, one finds that $\Delta_\ell$ are expressed in terms of $Q_{mf}, Q_{F}$, and $Q_{B}$,
\begin{align}
\Delta_{1}&= Q_{m1}/Q_{m2}, &\Delta_{2}&= Q_{m2}Q_{m3}Q_F, & \Delta_{3}&= Q_{m4}Q_{m6}Q_B, \cr
\Delta_{4}&= Q_{m5}/Q_{m6}, &\Delta_{5}&= Q_{m6}Q_{m7}Q_F, & \Delta_{6}&= Q_{m8}Q_{m2}Q_B.
\end{align}
This leads to the period \eqref{period} being the instanton factor squared:
\begin{align}
d=\Big(\prod^8_{f=1} Q_{mf}\Big) Q_F^2Q_B^2 = \mathfrak{q}^2.
\end{align}
In a similar fashion, all the $Q_{1(\ell)}$ are expressed in terms of $y_i, A,$ and $\mathfrak{q}$, and so are all K\"ahler parameters. 
The results for the other parameters are summarized in  Appendix \ref{app:Kahler}.
Since it is known that the instanton factor of the 5d $Sp(1)$ gauge theory with $N_f=8$ flavor is identified 
as the modulus of the compactified torus of the E-string,
we find that the period of our Tao diagram is also given by this modulus.
This is consistent with the intuition that the spiral structure will correspond to the KK mode of the E-string
and, thus, the cyclic structure of the spiral in our Tao diagram is a key to the uplift to 6d. 
In the next section, we compute the partition function
and give quantitative support for this claim.

Here, we briefly comment on the subtle difference between the parameters used in
the Nekrasov partition function for the 5d $Sp(1)$ gauge theory with eight flavors and
those used in the E-string partition function, which is clarified in \cite{Kim:2014dza}.
Our parameters $y_f$, $\mathfrak{q}$, and $A$ are those for the 5d Nekrasov partition function.
If we would like to obtain the E-string partition function in an $\hat{E}_8$ manifest form,
we need to use other parameters $y'_8$ and $A'$ defined as
\begin{align}
y'_8 = y_8 \,\mathfrak{q}^{-2}, \qquad A' = A\, \mathfrak{q}\, y_8^{-1} ,
\label{eq:Estring-param}
\end{align}
while the other parameters are identical. 
Since our formula for the partition function will be given in a closed form and in a way that does not depend on the detail of the
parametrization of the K\"ahler parameters,
it should be possible to interpret it either as the 5d Nekrasov partition function 
or as the E-string partition function depending on which parametrization we use. 
However, when we compare our formula with the known result \cite{Hwang:2014uwa,Kim:2014dza},
we use the parameters for the 5d Nekrasov partition function.

\section{E-string partition function via topological vertex}\label{sec:E-stringviaTopver}
In the previous section, we found a new web description of the 6d E-string theory. 
In this section,
we compute the E-string partition function by applying
the topological vertex computation to our web.
We will see that our partition function precisely agrees with
the partition function computed by a completely different method \cite{Hwang:2014uwa,Kim:2014dza}.
For simplicity, we concentrate on the 
self-dual $\Omega$-background $\varepsilon_1= - \varepsilon_2$ in this paper.

 \begin{figure}[t]
 \begin{center}
\includegraphics[width=15cm, bb=0 0 917 191]{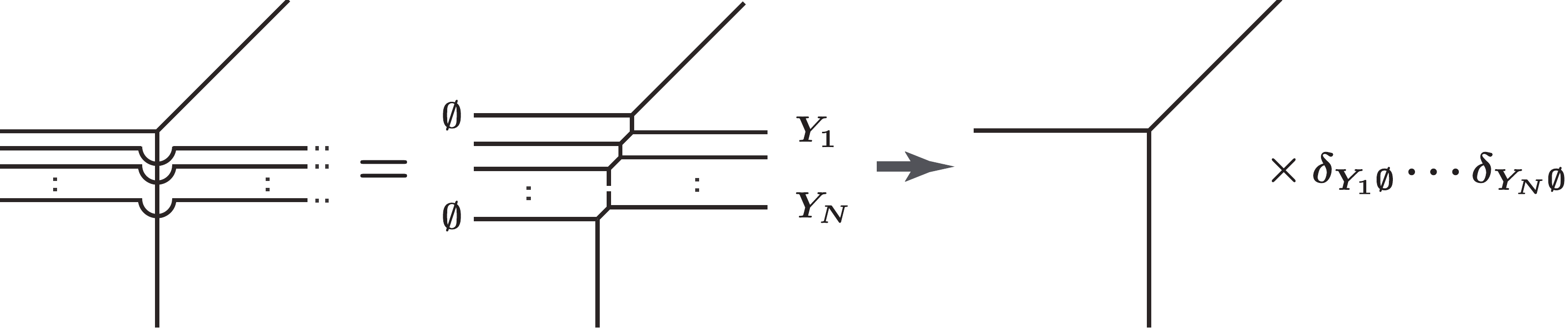}
 \end{center}
 \caption{\small In topological vertex formalism, jumping 5-branes are actually decoupled.}
 \label{fig;deg}
\end{figure}
\subsection{Combinatorial expression of E-string partition function}

In this section, we compute the topological string partition function of the Tao web Figure \ref{fig:taoweb}.
This Tao web contains a number of 5-branes jumping other 5-branes as we discussed in the previous section.
On the left-hand side of Figure \ref{fig;deg}, such a configuration is illustrated.
This jumping is realized by degenerating the corresponding K\"ahler parameters as in the middle diagram in the topological vertex formalism,
and these jumping 5-branes are decoupled from the nontrivial trivalent vertex as in the right hand of Figure \ref{fig;deg}.
We therefore need to take only this nontrivial vertex into account in the topological vertex computation.

\begin{figure}[t]
 \begin{center}
\includegraphics[width=6.5cm, bb=0 0 744 584]{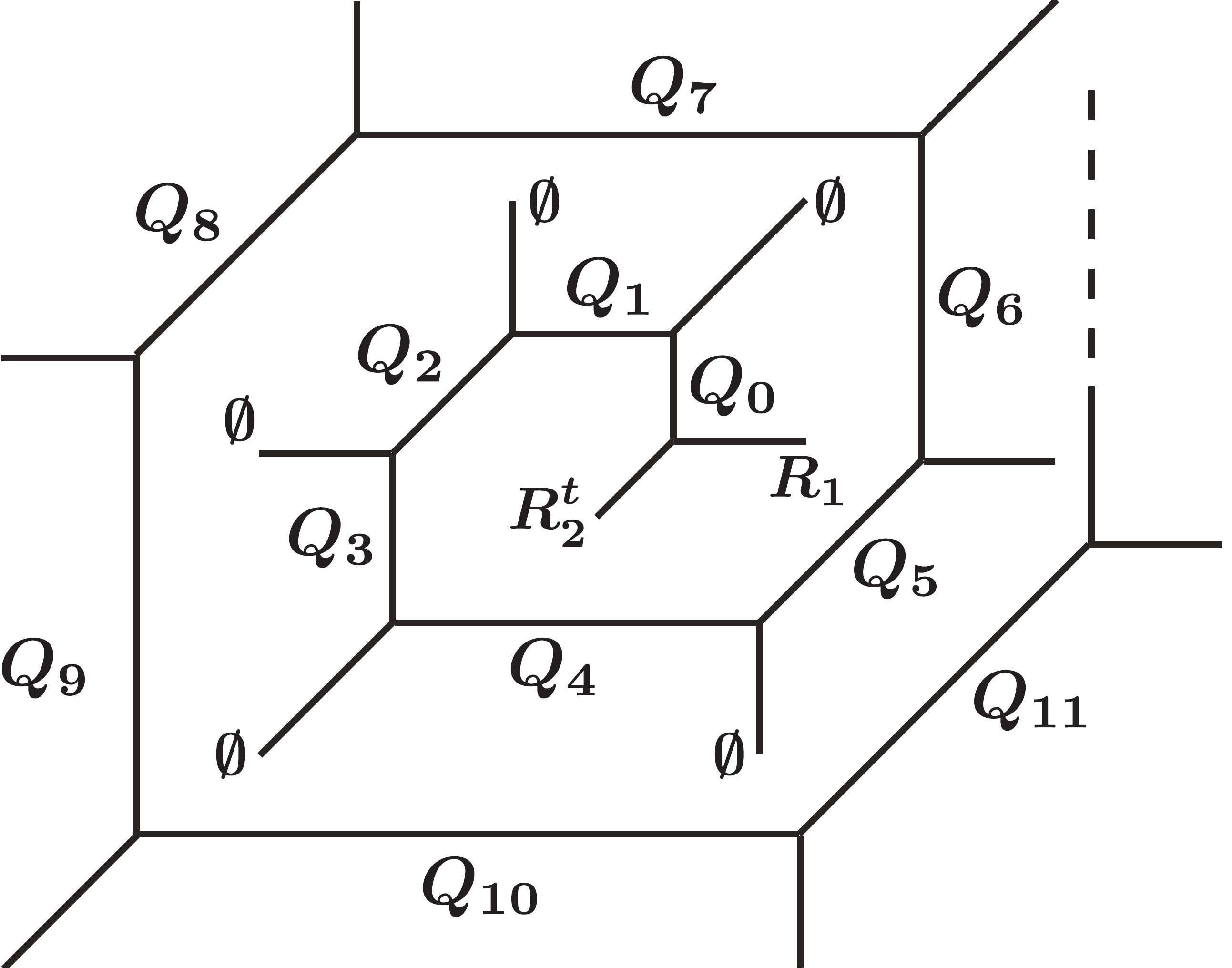}
 \end{center}
 \caption{\small A spiral leg which is a building block of the Tao web.}
 \label{fig:spiralleg}
\end{figure}
In the topological vertex computation of Nekrasov partition functions,
we first decompose a toric web into basic building blocks \cite{Iqbal:2003ix,Iqbal:2003zz,Eguchi:2003sj}.
Following this idea, we consider the basic spiral block Figure \ref{fig:spiralleg} of the Tao diagram.
On the two internal edges located at the starting point of the spiral,
two generic Young diagrams $R_1$ and $R_2^t$ are assigned.
This is because we need to glue six such building blocks 
for reconstructing the original Tao diagram in the topological vertex formalism.
This gluing procedure is performed by summation over all such Young diagrams associated with the legs we want to glue together.
All the other external legs without labels are associated with the empty Young diagram $\emptyset$.
The partition function for this sub-diagram is then given by the topological vertex as
\begin{align}
\label{block}
Z^{\, leg}_{R_1,R_2}(Q_0;\,Q_1,Q_2,Q_3,\ldots;q)
=\sum_{Y_0,Y_1,Y_2,\ldots}
C_{R_1R_2^tY_0}(q)
\prod_{k=0}^\infty\,(-Q_k)^{\vert Y_k\vert}\,
C_{\emptyset  Y_{k}^tY_{k+1}}(q).
\end{align}
In the original Tao diagram, the exponentiated K\"ahler parameters $Q_i$ are strongly correlated with each other because of the 7-brane construction of the web,
but we temporarily assign generic values to all these K\"ahler parameters just for simplicity. 
By substituting the definition of the topological vertex function $C_{R_1R_2R_3}(q)$,
we can write down this sub-diagram explicitly in terms of Shur and skew-Shur functions.
The result is
\begin{align}
\label{block2}
Z^{\, leg}_{R_1,R_2}(Q_0;\,Q_1,Q_2,Q_3,\cdots;q)
=q^{\frac{\kappa_{R_1}}{2}}\sum_{Y,Y_0,Y_1,Y_2,\cdots}&
S_{Y_0}(q^{\rho})S_{R_1^t/Y}(q^{Y_0+\rho})
S_{R_2^t/Y}(q^{Y_0^t+\rho})\nonumber\\
\times \prod_{k=0}^\infty&\,(-Q_k)^{\vert Y_k\vert}\,
S_{Y_{k+1}}(q^{\rho})S_{Y_k^t}(q^{Y_{k+1}^t+\rho})
.
\end{align}
Notice that the topological vertex function has the cyclic symmetry $C_{R_1R_2R_3}=C_{R_2R_3R_1}=C_{R_3R_1R_2}$,
and thus the above expression in terms of Schur functions is not unique.
Finding infinite product expressions of the sub-diagram by performing all the summations is not easy,
and it is an interesting open problem.
Instead of solving it,
we deal with this sub-diagram as a series expansion in the K\"ahler parameters in a later subsection. 
We can then evaluate this building block up to any order 
as far as a computer runs.

The generating function of the elliptic genera of the E-strings,
which is the topological string partition function for the local $\frac{1}{2}$K3,
therefore takes the following combinatorial form
\begin{align}
\label{MasterFormula}
Z_{E{\textrm{-}}string\,\,Tao}(y,d,A;q)
=M(q)\sum_{R_1,\ldots,R_6}
\prod_{\ell=1}^6
(-I_{\ell})^{\vert R_{\ell} \vert}\,
Z^{\, leg}_{R_{\ell},R_{\ell+1}}(Q_{0(\ell)};\,Q_{1(\ell)},Q_{2(\ell)},Q_{3(\ell)},\cdots;q),
\end{align}
where $R_7:=R_1$ and $Q_{0(\ell)}$s are
\begin{align}
Q_{0(1)}=Q_{m1},\,\,
Q_{0(2)}=Q_{m3},\,\,
Q_{0(3)}=Q_{m4},\,\,
Q_{0(4)}=Q_{m5},\,\,
Q_{0(5)}=Q_{m7},\,\,
Q_{0(6)}=Q_{m8}.
\end{align}
Here we have added the contribution from the constant map by hand which takes the form of the MacMahon function
\begin{align}
M(q)={\rm PE}\Big[\frac{q}{(1-q)^2}\Big],
\end{align}
where the Plethystic exponential PE is defined as
\begin{align}
 {\rm PE}\Big[f(\cdot)\Big] = {\rm exp}\Big[\sum_{n=1}^\infty \frac{1}{n} f(\cdot ^n)\Big].
\end{align}
The K\"ahler parameters $I_\ell$ for the six sides of the central hexagon in Figure \ref{fig:taoweb} are:
\begin{align}
I_1=Q_BQ_{m2},\,\,
I_2=Q_{m2}^{-1},\,\,
I_3=Q_FQ_{m2},\,\,
I_4=Q_BQ_{m6},\,\,
I_5=Q_{m6}^{-1},\,\,
I_6=Q_FQ_{m6}.
\end{align}
This is a simple and closed expression for the generating function of the E-string elliptic genera.
By expanding it in terms of the Coulomb branch parameter $A$,
we can obtain the $n$ E-string elliptic genus as the coefficient of $A^n$.

Here, we comment on the discrete symmetry that our partition function enjoys.
It is straightforward to see that the expression (\ref{MasterFormula}) is invariant under the following transformation
\begin{eqnarray}
Q_{i (\ell)} \to Q_{i (\ell+1)},\qquad
I_{\ell} \to I_{\ell+1},
\label{eq:Rotation}
\end{eqnarray}
where $Q_{i (7)} = Q_{i (1)}$ and $I_7=I_1$.
This transformation is generated by the ``$\pi / 3$ rotation'' of the Tao diagram in Figure \ref{fig:notation},
which transforms one of the arms to the next one.
By using the explicit parametrization of the  K\"ahler moduli parameters summarized in Appendix \ref{app:Kahler},
the transformation in (\ref{eq:Rotation}) is rewritten in terms of the mass parameters $y_i$ and the instanton factor $\mathfrak{q}$ as
\begin{eqnarray}
y_f \to \frac{\mathfrak{q}^{\frac{1}{2}} y_f}{\prod_{i=1}^8 \left( y_i \right) ^{\frac{1}{4}}},
\qquad
A \to \frac{\mathfrak{q}^{\frac{1}{2}} A}{\prod_{i=1}^8 \left( y_i \right) ^{\frac{1}{4}}},
\label{eq:PartofE9}
\end{eqnarray}
combined with a certain $SO(16)$ Weyl transformation%
\footnote{
More concretely, this $SO(16)$ Weyl transformation is given by
\begin{eqnarray*}
y_2 \leftrightarrow y_6{}^{-1},
\qquad
y_1 \to y_3 \to y_4 \to y_5 \to y_7 \to y_8 \to y_1.
\end{eqnarray*}
}
acting on $y_i$ ($i=1,\ldots 8$.) 
Or, if we use the parameters for the E-string \eqref{eq:Estring-param}, it is rewritten as
\begin{eqnarray}
y'_f \to \frac{y'_f}{\prod_{i=1}^8 \left( y'_i \right) ^{\frac{1}{4}}},
\qquad
A' \to A'.
\end{eqnarray}
This transformation 
can be interpreted as part of the expected $E_9$ symmetry.
This is analogous to the ``fiber-base duality map" studied in \cite{Bao:2011rc, Mitev:2014jza}.

\subsection{Comparing with the elliptic genus approach}

Now that we have the all-order generating function of the elliptic genera (\ref{MasterFormula})
that is the topological string partition function for the local $\frac{1}{2}$K3 in the 
self-dual $\Omega$-background,
we can derive various results.
One surprising application is reproducing the known partial results on E-strings from our partition function. The expansion of the generating function in $A$,
\begin{align}\label{Etaoextra}
Z_{E{\textrm{-}}string\,\,Tao}(y,\mathfrak{q},A)
=Z_{extra}(y,\mathfrak{q})\,M(q)\left(1+
\sum_{n=1}^\infty
A^n\,
Z_{n{\textrm{-}}string}(y,\mathfrak{q})\right),
\end{align}
should lead to the known elliptic genera $Z_{\,n{\textrm{-}}string}(y_i,\mathfrak{q};q)$ 
of the E-string theory \cite{Hosono:1999qc,Kim:2014dza}.
 In this expression we introduce the extra factor $Z_{extra}$ as the $A$-independent coefficient
which comes from the additional contribution that is not contained in the E-string theory. The E-string partition function is therefore defined by removing the extra contribution from the E-string Tao partition function \eqref{Etaoextra}.
The normalized partition function $Z_{E\textrm{-}string}$ corresponds to the E-string theory
\begin{align}\label{Estringtao}
Z_{E{\textrm{-}}string\,\,Tao}(y,\mathfrak{q},A)
=Z_{extra}(y,\mathfrak{q})\,Z_{E\textrm{-}string}(y,\mathfrak{q},A).
\end{align} 
The E-string partition function can be expressed as the plethystic exponential form \begin{align}
Z_{E\textrm{-}string} ={\rm PE}\bigg[\sum_{m=0}^{\infty}\mathcal{F}_m(y,A,q)\mathfrak{q}^m\bigg].
\end{align}
Likewise, the extra factor\footnote{The extra factor 
is an analogue of what is called the contribution coming from effective classes in \cite{KonishiMinabe}, the missing state/sector in \cite{Bergman:2013ala}, non full spin content in \cite{Hayashi:2013qwa}, U(1) factor in \cite{Bao:2013pwa}, and stringy contributions in \cite{Hwang:2014uwa}.}
 can also be expressed as
\begin{align}
Z_{\,extra}(y,\mathfrak{q})\,=
\textrm{PE}\bigg[
\sum_{k=0}^\infty \mathcal{E}_k(y,q)\,\mathfrak{q}^k
\bigg].
\end{align}
In what follows, we compute the E-string Tao partition function as the instanton expansion. At each order of the expansion we determine the extra factor and the E-string partition function
\begin{align}
{\rm PE}\Big[\big(\mathcal{E}_0+\mathcal{F}_0\big)+\big(\mathcal{E}_1+\mathcal{F}_1\big)\mathfrak{q}+\big(\mathcal{E}_2+\mathcal{F}_2\big)\mathfrak{q}^2+\cdots
\Big].
\end{align}


\subsection{Perturbative and instanton parts}

\begin{figure}[t]
 \begin{center}
\includegraphics[width=11cm, bb=0 0 1456 724]{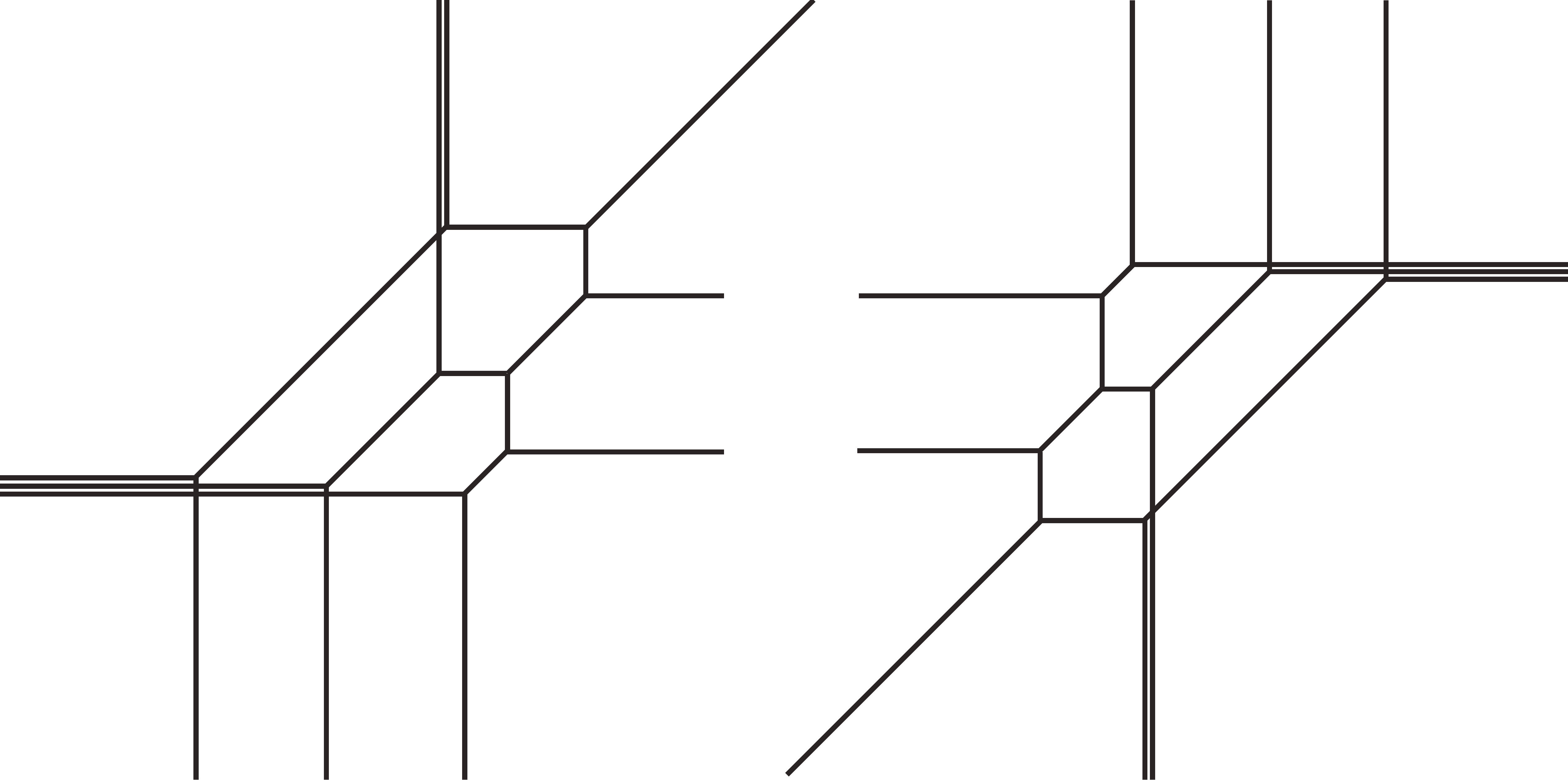}
 \end{center}
 \caption{\small Sub-diagrams of the Tao web that give the perturbative part.}
 \label{fig:pert}
\end{figure}
Now we expand the partition function in terms of the instanton factor and compare to
the 5d Nekrasov partition function or the E-string partition function
 studied in \cite{Hwang:2014uwa,Kim:2014dza}.
Although our Tao diagram spirally extends to infinity and thus \eqref{MasterFormula} includes infinitely many Young diagram sums, it turns out that we can truncate the arm at a finite place if we compute only the finite order of instanton expansion.\\

\noindent\underline{\bf Perturbative} \\
First, we start by computing the perturbative part.
In the topological vertex computation of 5d Nekrasov partition functions,
the perturbative contribution to a partition function is given by
the sub-diagrams that correspond to turning off the instanton factor $\mathfrak{q}$ \cite{Iqbal:2003ix,Iqbal:2003zz,Eguchi:2003sj}.
Indeed, we see that by 
taking into account the parametrization in Appendix \ref{app:Kahler}, setting $\mathfrak{q}=0$ makes most Young diagram summations in \eqref{MasterFormula} become trivial (only empty Young diagrams contribute). 
Figure \ref{fig:pert} shows two sub-pieces, which survive after setting $\mathfrak{q}=0$. %

In practice, it is more convenient to use the technique developed in \cite{Iqbal:2004ne} rather than directly using \eqref{MasterFormula}.
In this computation only the Young diagram sums for the horizontal lines corresponding to D5-branes remain.
We summed over these Young diagrams up to a total of ten boxes for each of the two sub-pieces, 
which corresponds to the expansion of the partition function in terms of $y_3$ and $y_7$ up to degree ten.

 The topological vertex method then leads to the following expression of the perturbative partition function:
\begin{eqnarray}
Z_{\, pert+extra}
&=&  {\rm PE}  \bigg[ \frac{q}{(1-q)^2}
\bigg( 1 - (y_1 + y_3 + y_4 + y_5 + y_7 + y_8) A^{-1}
\nonumber \\
&& \qquad
- ( y_1 + y_2 + y_2{}^{-1} + y_3 + y_4 + y_5 + y_6 + y_6{}^{-1} + y_7 + y_8  ) A + 2 A^2
\bigg)
\nonumber \\
&& \qquad
+ \mathcal{E}_0(y)
+ \mathcal{O} (y_3{}^{11}, y_7{}^{11})
\bigg]
\label{pert}
\end{eqnarray}
Taking into account that $y_3$ and $y_7$ to the power more than one do not appear up to 10, we expect that the higher-order correction $\mathcal{O} (y_3{}^{11}, y_7{}^{11})$ may also vanish.
Here, $\mathcal{E}_0(y)$ is the factor which does not depend on $A$ and is given by 
\begin{eqnarray}
\mathcal{E}_0(y) = \frac{q}{(1-q)^2}\left( y_1 y_2{}^{-1} + y_1 y_3 + y_2 y_3 + y_5 y_6{}^{-1} + y_5 y_7 + y_6 y_7\right).
\end{eqnarray}
We should regard such factor as the extra factor, and it should be removed by hand.
If we suitably use the analytic continuation with a certain regularization,
\begin{eqnarray}
{\rm PE}  \bigg[ \frac{q}{(1-q)^2}  Q \bigg] 
\to {\rm PE}  \bigg[ \frac{q}{(1-q)^2 } Q^{-1} \bigg],
\end{eqnarray}
the obtained result is consistent with the expected perturbative contribution calculated in \cite{Kim:2014dza}:
\begin{eqnarray}\label{pertanaly}
Z_{\, pert}=
{\rm PE} \big[\mathcal{F}_0(y)\big]
=  {\rm PE}  \bigg[ \frac{q}{(1-q)^2} 
\left(1 - \chi_{\bf 16}(y) A + 2 A^2 \right)
\bigg],
\end{eqnarray}
where we have defined the character of the $\bf16$ representation of $SO(16)$ flavor symmetry
\begin{eqnarray}
\chi_{\bf 16}(y) = \sum_{i=1}^{8} ( y_i + y_i{}^{-1} ).
\end{eqnarray}
Such an analytic continuation usually 
appears in the flop transition \cite{Iqbal:2004ne,Konishi:2006ev,Taki:2008hb} 
of topological strings on toric Calabi--Yau.
Discussing flop transition of a Tao web seriously is an interesting open problem,
but in practice we use the above continuation as a mathematical trick.

It should be emphasized that when we define the instanton contribution, we should divide  
the E-string Tao partition function by the perturbative part (\ref{pert}) for which the analytic continuation is not performed:
 \begin{eqnarray}
 Z_{\, inst+extra} = \frac{Z_{E{\textrm{-}}string\,\,Tao} }{Z_{\,pert+extra}}.
 \end{eqnarray}

\begin{figure}[t]
 \begin{center}
\includegraphics[width=8cm, bb=0 0 2500 1816]{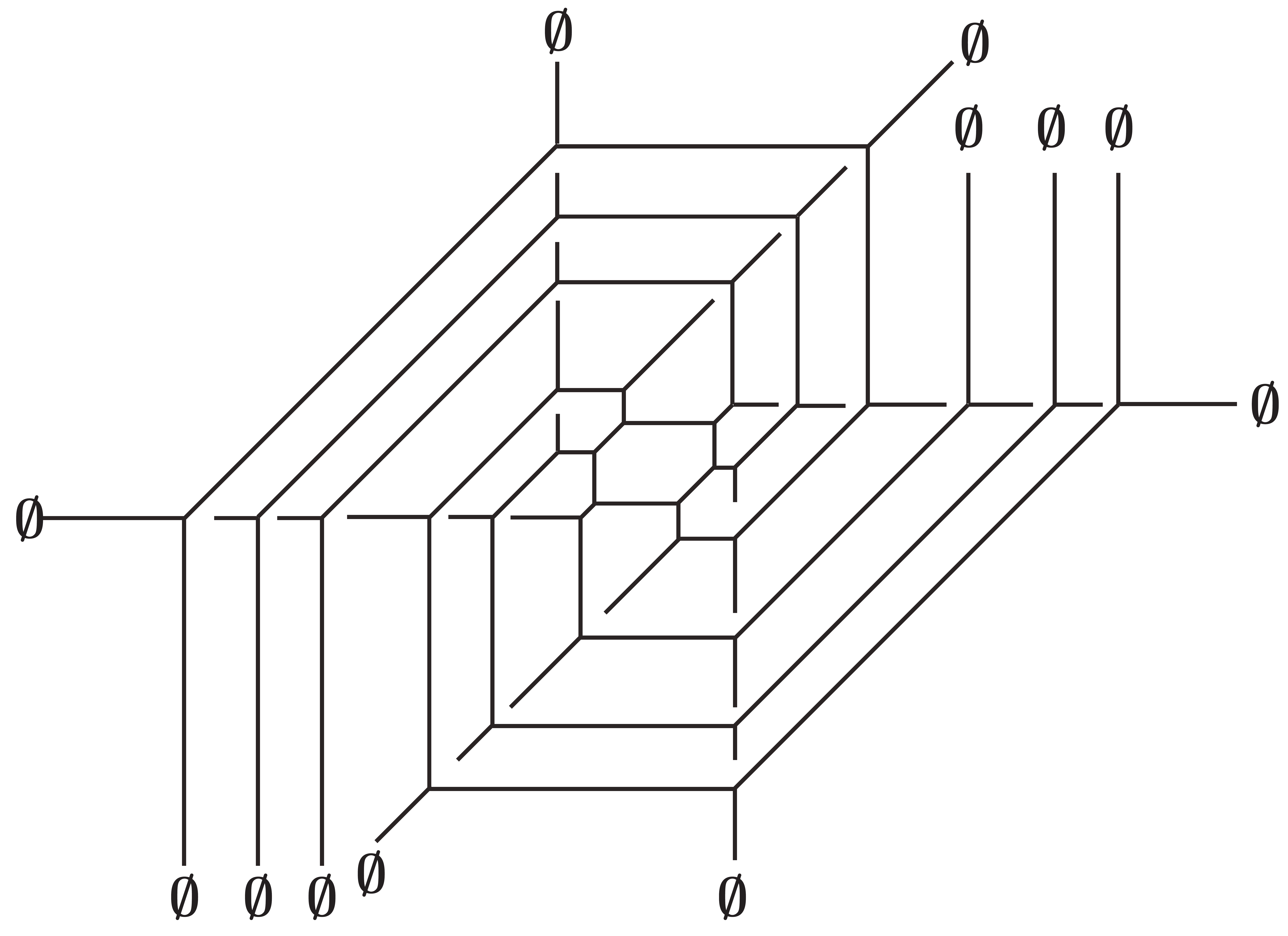}
 \end{center}
 \caption{\small Sub-diagram of the Tao web that gives the one-instanton part.}
 \label{fig:1-inst}
\end{figure}
\noindent\underline{\bf One-instanton}\\
When we compute the one-instanton contribution,
the K\"ahler parameters including high powers of instanton factor $\mathfrak{q}$ should all be truncated. When we consider the sub-diagram which includes instanton factor with power up to 1, 
we obtain the sub-diagram Figure \ref{fig:1-inst}.
Using {\textit{Mathematica}}, 
we show that the one-instanton contribution is a finite polynomial and given by 
\begin{align}
Z_{\, inst+extra} 
&= 1 
+ \mathfrak{q}\,\Bigg[ \frac{q}{(1-q)^2} 
\frac{A}{(1-A^2)^2}
\left( 
- (1+A^2) \chi_{\bf\overline{128}} (y) 
+ 2A \chi_{\bf128} (y)
\right) \cr
&\qquad\qquad + \mathcal{E}_1(y)  +\mathcal{O}( y_3{}^\frac{5}{2},y_7{}^{\frac{5}{2}})
\Bigg] +\mathcal{O} \big( \mathfrak{q}^2 \big)
\cr
&= \mathrm{PE} \Big[\big(\mathcal{E}_1 + \mathcal{F}_1 + \mathcal{O}( y_3{}^\frac{5}{2},y_7{}^{\frac{5}{2}})
\big)\mathfrak{q}\Big]
 + \mathcal{O} \bigl( \mathfrak{q}^2 \bigr),
\end{align}
$\mathcal{E}_1(y)$ is the factor which does not depend on $A$ given as
\begin{eqnarray}
\mathcal{E}_1(y) &=& \frac{q}{(1-q)^2}
\frac{1}{\prod_{i=1}^8 \left( y_i \right)^{\frac{1}{2}}}
(y_4  y_5 +y_1  y_3  y_4  y_5 +y_2  y_3  y_4  y_5 +y_4  y_6 +y_1  y_3  y_4  y_6 +y_2  y_3  y_4  y_6
\nonumber \\
&& \qquad \qquad
 +y_1  y_4  y_5  y_6 
+y_4  y_5  y_6  y_7 +y_1  y_3  y_4  y_5  y_6  y_7 +y_2  y_3  y_4  y_5  y_6  y_7 +y_1  y_8 
\nonumber \\
&& \qquad \qquad 
+y_2  y_8 +y_1  y_2  y_3  y_8  
+y_1  y_2  y_5  y_8 +y_1  y_5  y_7  y_8 +y_2  y_5  y_7  y_8 
\nonumber \\
&& \qquad \qquad 
+y_1  y_2  y_3  y_5  y_7  y_8  +y_1  y_6  y_7  y_8 
+y_2  y_6  y_7  y_8  + y_1  y_2  y_3  y_6  y_7  y_8 ).
\end{eqnarray}
The one-instanton computation result is given by
\begin{eqnarray}
\mathcal{F}_1
=\frac{q}{(1-q)^2} 
\left[
\frac{A}{(1-A^2)^2}
\left( 
- (1+A^2) \chi_{\bf\overline{128}} (y) 
+ 2A \chi_{\bf128} (y)
\right) 
\right],
\end{eqnarray}
or
\begin{eqnarray}\label{F1}
\mathcal{F}_1
= - \frac{q A}{2 (1-q)^2 (1+A)^2} 
(\chi_{\bf s} (y) + \chi_{ \bf c} (y))
+ \Big[ \{ y_i, A\} \to \{ - y_i , -A \} \Big].
\end{eqnarray}
In \eqref{F1} we used the following notation: The characters $\chi^{(n)}$  ($n\le 6$) are $n$th anti-symmetric tensor representations whose Dynkin label is given by $[0, 0, \ldots, 1, \ldots, 0 ]$ where each entry is zero except for the $n$th entry being one. One can also call them the characters of the fundamental weights of $SO(16)$ together with two spinor representations identified as $\chi_{\bf c}=\chi^{(7)}, \chi_{\bf s}=\chi^{(8)}$. 
We note that $\chi^{(2)}$, $\chi^{(4)}$, $\chi^{(6)}$, and $\chi^{(8)}$ are invariant under the transformation $y_i \to - y_i$, while $\chi^{(1)}$, $\chi^{(3)}$, $\chi^{(5)}$, and $\chi^{(7)}$ obtain a minus sign by this transformation.\footnote{In terms of the irreducible representations, the fundamental weights of $SO(16)$ are as follows: 
\begin{align*}
\chi^{(1)}&=\chi^{SO(16)}_{\bf 16}, &\chi^{(2)}&=\chi^{SO(16)}_{\bf 120},&\chi^{(3)}&=\chi^{SO(16)}_{\bf 560},&
\chi^{(4)}&=\chi^{SO(16)}_{\bf 1820},\\
\chi^{(5)}&=\chi^{SO(16)}_{\bf 4368},&\chi^{(6)}&=\chi^{SO(16)}_{\bf 8008}, &\chi^{(7)}&=\chi^{SO(16)}_{\bf\overline{128}},&\chi^{(8)}&=\chi^{SO(16)}_{\bf{128}}.
\end{align*}
}
From here on, we assume that the higher-order correction $\mathcal{O}( y_3{}^\frac{5}{2},y_7{}^{\frac{5}{2}})$ vanishes.

\noindent\underline{\bf Two-instanton}\\
The two-instanton contribution including the extra factor is 
\begin{eqnarray}
\mathrm{PE} \big[ (\mathcal{E}_2 + \mathcal{F}_2 + \mathcal{O}( y_3{}^\frac{5}{2},y_7{}^{\frac{5}{2}}) )\mathfrak{q}^2 \big],
\end{eqnarray}
where the extra factor $\mathcal{E}_2 $ 
is given by
\begin{align}
\mathcal{E}_2 =
 \frac{q}{(1-q)^2}
\bigg(&q+4+\frac{1}{q}+ \frac{y_1}{y_2} + \frac{y_2}{y_1} + \frac{1}{y_1 y_3} + \frac{1}{y_2 y_3} + 
 y_1 y_3 + y_2 y_3 + \frac{y_5}{y_6} + \frac{y_6}{y_5} \nonumber\\
 & +\frac{1}{y_5 y_7} + \frac{1}{ y_6 y_7 } + y_5 y_7 + y_6 y_7 \bigg).
\end{align}
The two-instanton contribution takes the following form:
\begin{eqnarray}
\mathcal{F}_2 = 
\frac{q}{(1-q)^2} \frac{1}{(1-q^{-1}A^2 )^2 ( 1- q A^2 )^2}
& \Bigl( c_1\, \chi^{(1)} + c_2\, \chi^{(2)} + c_3 \,\chi^{(3)} + c_4 \,\chi^{(4)} + c_5\, \chi^{(5)}
\nonumber\\
& + c_6\, \chi^{(6)} 
+ c_7 \,\chi_{\bf s}\,\chi_{\bf c} 
+ c_8\, (\chi_{\bf s}^2 + \chi_{\bf c}^2) + c_9 \Bigr).
\nonumber
\end{eqnarray}
The coefficient functions $c_{k}$ are functions of $q$ and $A$ such that they are expressed in terms of the $SU(2)$ characters of $q$, $\chi_{\bf dim}(q)$,  
for instance, $\chi_{\bf 2} (q) = q^{-1} + q$, $\chi_{\bf 3} (q) = q^{-2} +  1 + q^2$ and $\chi_{\bf 4} (q) = q^{-3} + q^{-1} + q + q^3$.
The coefficient functions $c_k$ are given as follows:
\begin{flalign}        
 c_1 &= - A (1 + A^2) \Bigl( (2 {\chi_{\bf 2} (q)}+1) (1+A^4)+ (2 {\chi_{\bf 2} (q)}+{\chi_{\bf 3} (q)}+1) A^2 \Bigr), &&
 \end{flalign}
\vspace{-10mm}
\begin{flalign}        
 c_2 &= \frac{A^2 ( 1 + A^2 + A^4 ) }{(1 + A^2)^2}
 \Bigl( (3 {\chi_{\bf 2} (q)}+2)(1+A^4) +(7 {\chi_{\bf 2} (q)}+6) A^2 \Bigr), &&
\end{flalign}
 \vspace{-10mm}
\begin{flalign}        
 c_3 &= - A (1 + A^2) \Bigl( (1+A^4) + ( 2 \chi_{\bf 3} (q) + 1) A^2 \Bigr), &&
\end{flalign}
\vspace{-10mm}
\begin{flalign}        
 c_4 &=  A^2 \Bigl( 2 (1+A^4)  +(2 {\chi_{\bf 2} (q)}+3) A^2 \Bigr),  &&
\end{flalign}
\vspace{-8mm}
\begin{flalign}        
 c_5 &= - 3 A^3 (1 + A^2) , &&
\end{flalign}
\vspace{-10mm}
\begin{flalign}        
 c_6 &= \frac{A^4 }{(1 + A^2)^2} \Bigl( 4 (1+A^4) - ({\chi_{\bf 2} (q)}-6) A^2 \Bigr), &&
\end{flalign}
\vspace{-8mm}
\begin{flalign}        
 c_7 &= - \frac{A^5 (1 + A^2)}{(1 - A^2)^4} \Bigl( 5 (1+A^4) - (2 \chi_{\bf 3} (q) + 4 ) A^2 \Bigr), &&
 \end{flalign}
\vspace{-8mm}
\begin{flalign}        
 c_8 &= \frac{A^6}{(1 - A^2)^4(1 + A^2)^2}  \Bigl( 6 (1+A^8) - ({\chi_{\bf 2} (q)}+3) A^2 (1 + A^4) - {\chi_{\bf 2} (q)} A^4 \Bigr), &&
 \end{flalign}
\vspace{-8mm}
\begin{flalign}        
 c_9 &= A^2 \Bigl( 4  ({\chi_{\bf 2} (q)}+{\chi_{\bf 3} (q)}+1) (1+A^4)  + (-2 {\chi_{\bf 2} (q)}+2 {\chi_{\bf 3} (q)}-{\chi_{\bf 4} (q)}+2) A^2 \Bigr). &&
\end{flalign}

\noindent\underline{\bf Three-instanton}\\
The three-instanton contribution including the extra factor is 
\begin{eqnarray}
\mathrm{PE} \big[ (\mathcal{E}_3 + \mathcal{F}_3 + \mathcal{O}( y_3{}^\frac{5}{2},y_7{}^{\frac{5}{2}}) )\mathfrak{q}^3 \big],
\end{eqnarray}
where the extra factor $\mathcal{E}_3$ is given by
\begin{eqnarray}
\mathcal{E}_3 = \mathcal{E}_1
- \frac{q}{(1-q)^2} 
\left( 
\sqrt{ \frac{y_1 y_4 y_5 y_6}{y_2 y_3 y_7 y_8} } + \sqrt{ \frac{y_1 y_2 y_5 y_8}{y_3 y_4 y_6 y_7} }
\right).
\end{eqnarray}
We do not have a clear understanding of why $\mathcal{E}_3 $ is very similar to $\mathcal{E}_1$ but slightly different.
The three-instanton contribution takes the following form:
\begin{align}
 \mathcal{F}_3=&~ \biggl[\frac{q ( \chi_{\bf s} + \chi_{\bf c} ) }%
{2(1 - q)^2 (1+A)^2(1 + q A)^2  (1 + q ^{-1} A)^2 (1 - q A^2)^2 (1 - q^{-1} A^2)^2 }
\nonumber \\
&\quad
\times \Bigl(  
  d_{1} \chi^{(1)} 
+ d_{2} \chi^{(2)} 
+ d_{3} \chi^{(3)} 
+ d_{4} \chi^{(4)}
+ d_{5} \chi^{(5)} 
+ d_{6} \chi^{(6)}
 \nonumber \\
&  \qquad \quad
+ d_{7} \chi_{\bf s} \chi_{\bf c}
+ d_{8}   (\chi_{\bf s}{}^2 - \chi_{\bf s} \chi_{\bf c} + \chi_{\bf c}{}^2 )
+ d_{9} 
\Bigr)\biggr]
\nonumber \\
& +
\Bigl[ \{ A,y_i \} \to  \{-A, -y_i \} \text{ of the above} \Bigr],
\end{align}
where the coefficient functions $d_k$ are 
given as follows:
\begin{flalign}
d_{1} = &-A \Bigl(  (1 + 
        A^{12}) - (\chi_{\bf 2} (q) - 2) (A + A^{11}) + (2 \chi_{\bf 2} (q) + 4) (A^2 + A^{10})\cr
        &\qquad+ (4 \chi_{\bf 3} (q) + 2 \chi_{\bf 2} (q) + 8) (A^3 + A^9)
         + (3 \chi_{\bf 3} (q) + 10 \chi_{\bf 2} (q) + 10) (A^4 + A^8) \cr
        &\qquad + (8 \chi_{\bf 3} (q) + 5 \chi_{\bf 2} (q) + 14) (A^5 + 
         A^7) + (4 \chi_{\bf 3} (q) + 12 \chi_{\bf 2} (q) + 12) A^6  \Bigr),&&
\end{flalign}
\begin{flalign}         
d_{2} = &~ A^2 (1 + A + A^2)
        \,\Bigl(  
     2 (1 + A^8) + ( 
       A + A^7) + (2 \chi_{\bf 2} (q) + 5) (A^2 + A^6)\cr 
        &\qquad  + (3 \chi_{\bf 3} (q) + 2 \chi_{\bf 2} (q) + 4) (A^3 +
          A^5) + (-\chi_{\bf 3} (q) + 4 \chi_{\bf 2} (q) + 9) A^4\Bigr),&&  
\end{flalign}
\begin{flalign}
d_{3} = & - A^3 \Bigl(  3 (1 + A^8) + (\chi_{\bf 2} (q) + 4)(A + A^7) + (2 \chi_{\bf 2} (q) + 12) (A^2 + A^6)\cr 
        &\qquad + (2 \chi_{\bf 3} (q) + 6 \chi_{\bf 2} (q) + 12) (A^3 + A^5) + (\chi_{\bf 3} (q) + 6 \chi_{\bf 2} (q) + 19) A^4 \Bigr), &&
\end{flalign}
\begin{flalign}         
d_{4} = &~ \frac{A^4 (1 + A^2)}{(1 - A + A^2)^2} \cr
&\times  
     \Bigl(4 (1 + A^8) + (2 \chi_{\bf 2} (q) - 3) (A + A^7)   - (2 \chi_{\bf 2} (q) - 14) (A^2 + A^6) \cr
     &\quad + (\chi_{\bf 3} (q) + 8 \chi_{\bf 2} (q) - 8) (A^3 +A^5) + (-2 \chi_{\bf 3} (q) - 8 \chi_{\bf 2} (q) + 18) A^4\Bigr),&&
\end{flalign}
\begin{flalign}         
d_{5} = &  - \frac{A^5}{(1 - A + A^2)^2 }\cr
        &\times\Bigl( 5 (1 + A^8) + (3 \chi_{\bf 2} (q) - 4) (A + A^7) - (4 \chi_{\bf 2} (q) - 16) (A^2 +A^6)\cr
        &\quad + (10 \chi_{\bf 2} (q) - 10) (A^3 + A^5) - (\chi_{\bf 3} (q) + 10 \chi_{\bf 2} (q) - 21) A^4\Bigr),&&
\end{flalign}
\begin{flalign}         
d_{6} = &~ \frac{A^6}{(1 - A + A^2)^2}\Bigl( 6 (1 + A^6) + (4 \chi_{\bf 2} (q) - 5) (A + A^5) \cr
       &\qquad+ (-6 \chi_{\bf 2} (q) + 12) (A^2 + 
         A^4) + (-\chi_{\bf 3} (q) + 8 \chi_{\bf 2} (q) - 7) A^3 \Bigr),&&
\end{flalign}
\begin{flalign}         
d_{7} = &-\frac{A^7}{(1 - A)^4 (1 + A)^4 }
\Bigl( 7 (1 + A^8) + 5 \chi_{\bf 2} (q) (A + A^7) - 
      4 \chi_{\bf 2} (q) (A^2 + A^6)\cr
      &\qquad - (2 \chi_{\bf 3} (q) + \chi_{\bf 2} (q) + 2) (A^3 + A^5) + 
      2 A^4\Bigr),&&
\end{flalign}
\begin{flalign}         
d_{8} =& ~\frac{A^8}{(1 - A + A^2)^2 (1 -  A)^4 (1 + A)^4  }
\Bigl(
      8 (1 + A^{10}) + (6 \chi_{\bf 2} (q) - 7) (A + A^9) 
 \cr
& \qquad   
      + (-10 \chi_{\bf 2} (q) + 6) (A^2 +  A^8) 
     - (3 \chi_{\bf 3} (q) - 6 \chi_{\bf 2} (q) + 3) (A^3 + A^7)\cr
     & \qquad
     + (2 \chi_{\bf 3} (q) - 2 \chi_{\bf 2} (q) + 4) (A^4 + A^6) 
     - (2 \chi_{\bf 3} (q) + 4) A^5 
 \Bigr), &&
\end{flalign}
\begin{flalign}         
d_{9} = & ~\frac{A}{(1 - A + A^2)^2}\cr
&\times\Bigl( -2 \chi_{\bf 2} (q) (1 + A^{16}) + 
      4 \chi_{\bf 2} (q) (A + A^{15}) + (4 \chi_{\bf 3} (q) - 4 \chi_{\bf 2} (q) + 3) (A^2 + A^{14}) \cr
      &\quad + 
      10 \chi_{\bf 2} (q) (A^3 + A^{13})
     + (2 \chi_{\bf 4} (q) + 6 \chi_{\bf 3} (q) - 6 \chi_{\bf 2} (q) + 4) (A^4 + 
         A^{12}) \cr
      &\quad - (6 \chi_{\bf 4} (q) - 6 \chi_{\bf 3} (q) - 14 \chi_{\bf 2} (q) - 6) (A^5 + A^{11})\cr
      &\quad  - (\chi_{\bf 5}(q) - 10 \chi_{\bf 4} (q) - 4 \chi_{\bf 3} (q) - 4 \chi_{\bf 2} (q) - 2) (A^6 + A^{10}) \cr
      &\quad+ (2 \chi_{\bf 5}(q) - 12 \chi_{\bf 4} (q) + 
         16 \chi_{\bf 3} (q) + 8 \chi_{\bf 2} (q) + 14) (A^7 + A^9)\cr
      &\quad    - (3 \chi_{\bf 5}(q) - 12 \chi_{\bf 4} (q) + 3 \chi_{\bf 3} (q) - 
         8 \chi_{\bf 2} (q) + 4) A^8\Bigr).&&
\end{flalign}

\noindent\underline{\bf Four-instanton}\\
The four-instanton contribution including the extra factor is 
\begin{eqnarray}
\mathrm{PE} \big[ (\mathcal{E}_4 + \mathcal{F}_4 + \mathcal{O}( y_3{}^\frac{5}{2},y_7{}^{\frac{5}{2}}) )\mathfrak{q}^4 \big],
\end{eqnarray}
where the extra factor $\mathcal{E}_4$ turned out to be identical to $\mathcal{E}_2$:
\begin{align}
\mathcal{E}_4 = \mathcal{E}_2.
\end{align}
The main part $\mathcal{F}_4$ is hard to write down explicitly.
Here, we write the expansion of $\mathcal{F}_4$ to order $A^2$:
\begin{align}
\mathcal{F}_4 = -
\Bigl( 
&
(3 \chi_{\bf 3} (q) + 4\chi_{\bf 2}(q) + 2) \chi^{(1)} 
+ 2 \chi_{\bf 2} (q) \chi^{(3)} + \chi^{(5)} + \chi^{(1)} \chi^{(2)}
\Bigr) A
\cr
+ \Bigl( 
& (5 \chi_{\bf 4} (q) + 6 \chi_{\bf 3} (q) + 11\chi_{\bf 2} (q)  + 8 ) \chi^{(2)} 
+ ( 4 \chi_{\bf 3} (q) + 4\chi_{\bf 2} (q) ) \chi^{(4)} 
+ (3 \chi_{\bf 2} (q) - 2) \chi^{(6)} 
 \cr
&
+ (4 \chi_{\bf 3}(q) + 3 \chi_{\bf 2}(q) + 2 ) ( \chi^{(1)} )^2
+  3 \chi_{\bf 2} (q) \chi^{(1)} \chi^{(3)}
+ 2 \chi^{(1)} \chi^{(5)} + 2 (\chi^{(2)} )^2  + 2 (\chi_{\bf s})^2 
\cr
&
+ ( 6 \chi_{\bf 5} (q) + 8 \chi_{\bf 4} (q) + 16 \chi_{\bf 3} (q) + 20 \chi_{\bf 2} (q) + 10 )
\Bigr) A^2 +\mathcal{O}(A^3).
\end{align}

To compare with the partition function computed in \cite{Hwang:2014uwa,Kim:2014dza}, we expand our result in terms of the Coulomb modulus $A$ (which corresponds to $w$ in \cite{Kim:2014dza}) and obtain
\begin{align}
Z_{E\textrm{-}string}={\rm PE}\Big[\sum_{m=0}^{\infty}\mathcal{F}_m(y,A,q)\mathfrak{q}^m\Big]={\rm PE}\Big[\frac{1}{(1-q)(1-q^{-1})}\sum_{n=1}^\infty \tilde{f}_n A^n\Big],
\end{align}
where
\begin{align}
 \tilde{f}_1&= \chi^{(1)} + \chi_{\bf c} \,\mathfrak{q} 
 +\Big( 2\chi_{\bf 2}(q)\chi^{(1)} + \chi^{(3)} + \chi^{(1)} \Big)\mathfrak{q}^2  + \Big(\chi^{(1)}\chi_{\bf s}+ 2\chi_{\bf 2}(q) \chi_{\bf c}   \Big)  \mathfrak{q}^3
\cr
& \quad 
+ \Bigl( 3 \chi_{\bf 3} (q) + 4\chi_{\bf 2}(q) + 2) \chi^{(1)} 
+ 2 \chi_{\bf 2} (q) \chi^{(3)} + \chi^{(5)} + \chi^{(1)} \chi^{(2)} \Bigr) \mathfrak{q}^4
+\mathcal{O}(\mathfrak{q}^5),\\
 \tilde{f}_2&=-  2- 2 \chi_{\bf s}\, \mathfrak{q}
-  \Big( 2 \chi^{(4)}+ (3\chi_{\bf 2}(q) + 2 )\chi^{(2)} +4(\chi_{\bf 3}(q) + \chi_{\bf 2}(q)+1) \Big)\mathfrak{q}^2 \nonumber\\
&\quad\, - \Big(2 \chi^{(2)}\chi_{\bf s}+3\chi_{\bf 2}(q) \chi^{(1)}\chi_{\bf c}+4(\chi_{\bf 3}(q)+\chi_{\bf 2}(q)+1)\chi_{\bf s}\Big)\mathfrak{q}^3
\cr
& \quad\, 
+ \Bigl( 
(5 \chi_{\bf 4} (q) + 6 \chi_{\bf 3} (q) + 11\chi_{\bf 2} (q)  + 8 ) \chi^{(2)} 
+ ( 4 \chi_{\bf 3} (q) + 4\chi_{\bf 2} (q) ) \chi^{(4)} 
+ (3 \chi_{\bf 2} (q) - 2) \chi^{(6)} 
 \cr
& \qquad \qquad
+ (4 \chi_{\bf 3}(q) + 3 \chi_{\bf 2}(q) + 2 ) ( \chi^{(1)} )^2
+  3 \chi_{\bf 2} (q) \chi^{(1)} \chi^{(3)}
+ 2 \chi^{(1)} \chi^{(5)} + 2 (\chi^{(2)} )^2  + 2 (\chi_{\bf s})^2 
\cr
& \qquad \qquad
+ ( 6 \chi_{\bf 5} (q) + 8 \chi_{\bf 4} (q) + 16 \chi_{\bf 3} (q) + 20 \chi_{\bf 2} (q) + 10 )
\Bigr) \mathfrak{q}^4
 +\mathcal{O}(\mathfrak{q}^5).
\end{align} 
Here, $\tilde{f}_1$ and $\tilde{f}_2$ are precisely those used in \cite{Kim:2014dza}.
We further checked the five-instanton contribution by inputing $y_i$ with some specific values and found that $\mathcal{F}_5$ is also consistent.\footnote{As an example, we substituted $y_1=1$, $y_2=2$, $y_3=1$, $y_4=3$, $y_5=1$, $y_6=2$, $y_7=1$, $y_8=3$, $q=5$. We did not consider simpler massless case because zero appears in the numerator at the middle of the computation if we substitute $y_1 = y_2 = \cdots = y_8 =1$ from the beginning.}
Hence, the result obtained from the topological vertex amplitude based on the Tao diagram agrees with the result obtained from the instanton calculus as well as elliptic genus of E-strings in \cite{Hwang:2014uwa,Kim:2014dza}.\footnote{
Taking into account our setup $\epsilon_1+\epsilon_2=0$ and the notation convention introduced earlier, one easily finds that $t$ and $u$ in \cite{Kim:2014dza} correspond to $1$ and $q$, respectively, in our convention. The higher dimensional irreducible representations used in \cite{Hwang:2014uwa,Kim:2014dza} can be expressed in terms of the fundamental weights. For example,  $~\chi^{SO(16)}_{\bf 1920}= \chi^{(1)}\chi_{\bf c} - \chi_{\bf s}, \quad 
\chi^{SO(16)}_{\bf \overline{1920}}= \chi^{(1)}\chi_{\bf s} - \chi_{\bf c},\quad
\chi^{SO(16)}_{\bf 13312}=\chi^{(2)}\chi_{\bf s}-  \chi^{(1)}\chi_{\bf c}$.}

\section{Discussion}\label{sec:discussion}
\begin{figure}[t]
 \begin{center}
\includegraphics[width=13cm, bb=0 0 1089 513]{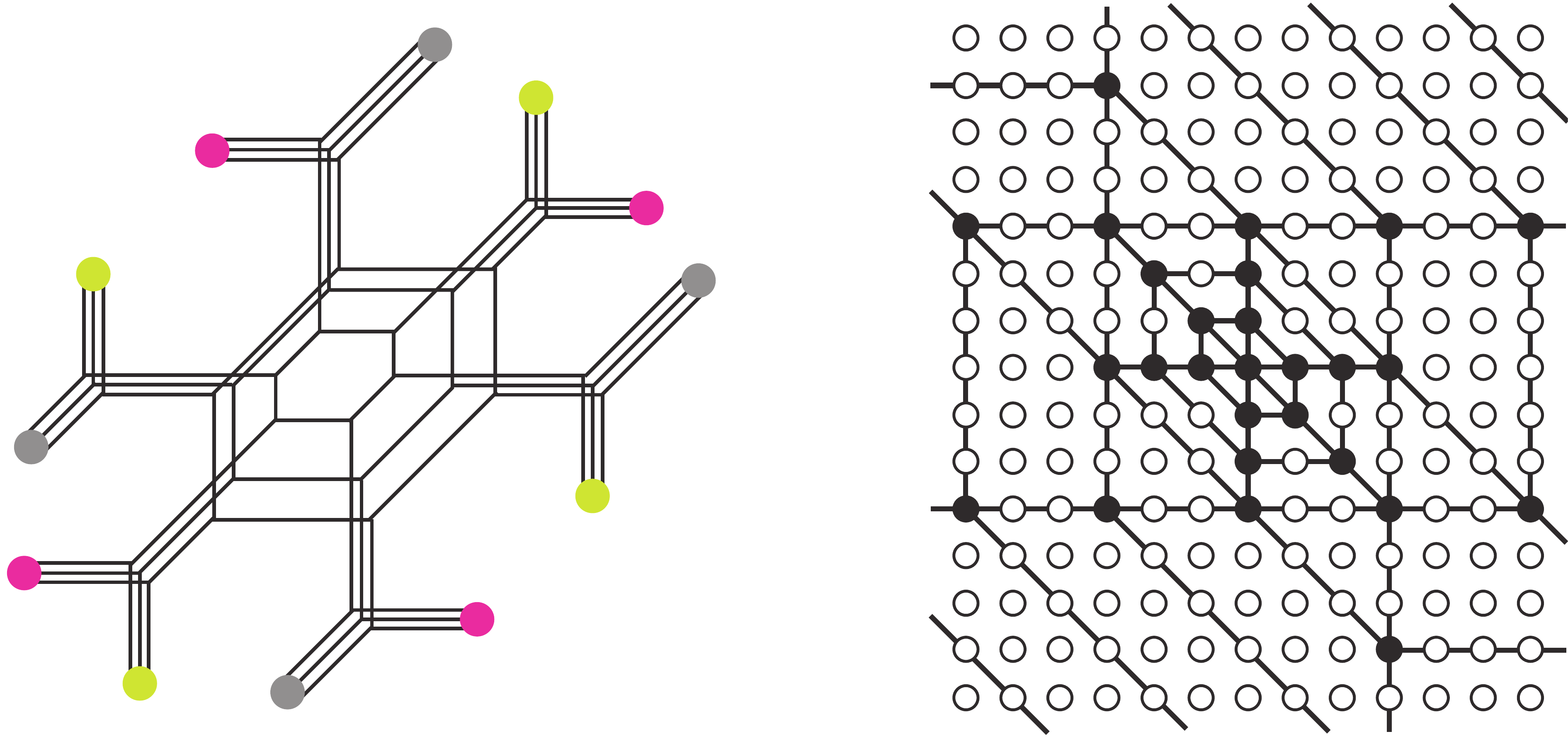}
 \end{center}
 \caption{\small A web diagram of class $\mathcal{T}$ for a higher-rank generalization of E-string theory.}
 \label{fig:HREST}
\end{figure}
In this article, we computed the partition function for 5d $SU(2)$ gauge theory with eight flavors based on a Tao diagram which is a $(p,q)$ 5-brane web that has spirally rotating arms with the period accounting for the instanton configuration of the theory. As shown  
in \cite{Kim:2014dza}, the Nekrasov function coincides with the elliptic genera of the E-string theory.  
Applying the topological vertex to the Tao web diagram, we reproduced the same partition function as \cite{Kim:2014dza} up to four instantons. Thus this gives a new way of studying 6d $(1,0)$ theories via Tao diagrams with type IIB perspectives. We note that there are extra factors in our computation which do not depend on the Coulomb modulus which we mode out by hand to reproduce the correct partition function.

Although we have considered a simple Tao diagram like Figure \ref{fig:taoweb} in the main text,
various Tao-type diagrams are possible which are of spiral webs with cyclic structure. We propose to call a collection of Tao diagrams
 ``class $\mathcal{T}$\hspace{-0.5mm}(ao).'' 

A typical example of theories of class $\mathcal{T}$ would be the higher-rank E-string theory.
An easy way to find such a web would be exploring consistent black--white grid (or toric-like) diagrams with spiral belts in the triangulated diagram. For instance, see Figure \ref{fig:HREST} which may describe 5d $Sp(N)$ gauge theory with a massless antisymmetric hypermultiplet.
The diagram in Figure \ref{fig:HREST} has $N$ normalizable deformations, namely $N$-dimensional Coulomb branch,
as there are $N$ internal points in the toric-like diagram.
Moreover, the web is a collection of $N$ coinciding E-string Tao webs 
that are essentially decoupled\footnote{This situation is same as the web for the higher-rank $E_n$ theory in \cite{Benini:2009gi,Kim:2014nqa}, which is the case where the antisymmetric hypermultiplet is massless. It is worth noting that the case with massive antisymmetric hypermultiplet is not described by Figure \ref{fig:HREST}. } 
from each other in this web realization.
This E-string-like diagram with multidimensional Coulomb branch like Figure \ref{fig:HREST}
 is a candidate for 5-brane web description of the rank-$N$ E-string theory.

\begin{figure}[t]
 \begin{center}
\includegraphics[width=14cm, bb=0 0 2478 1113]{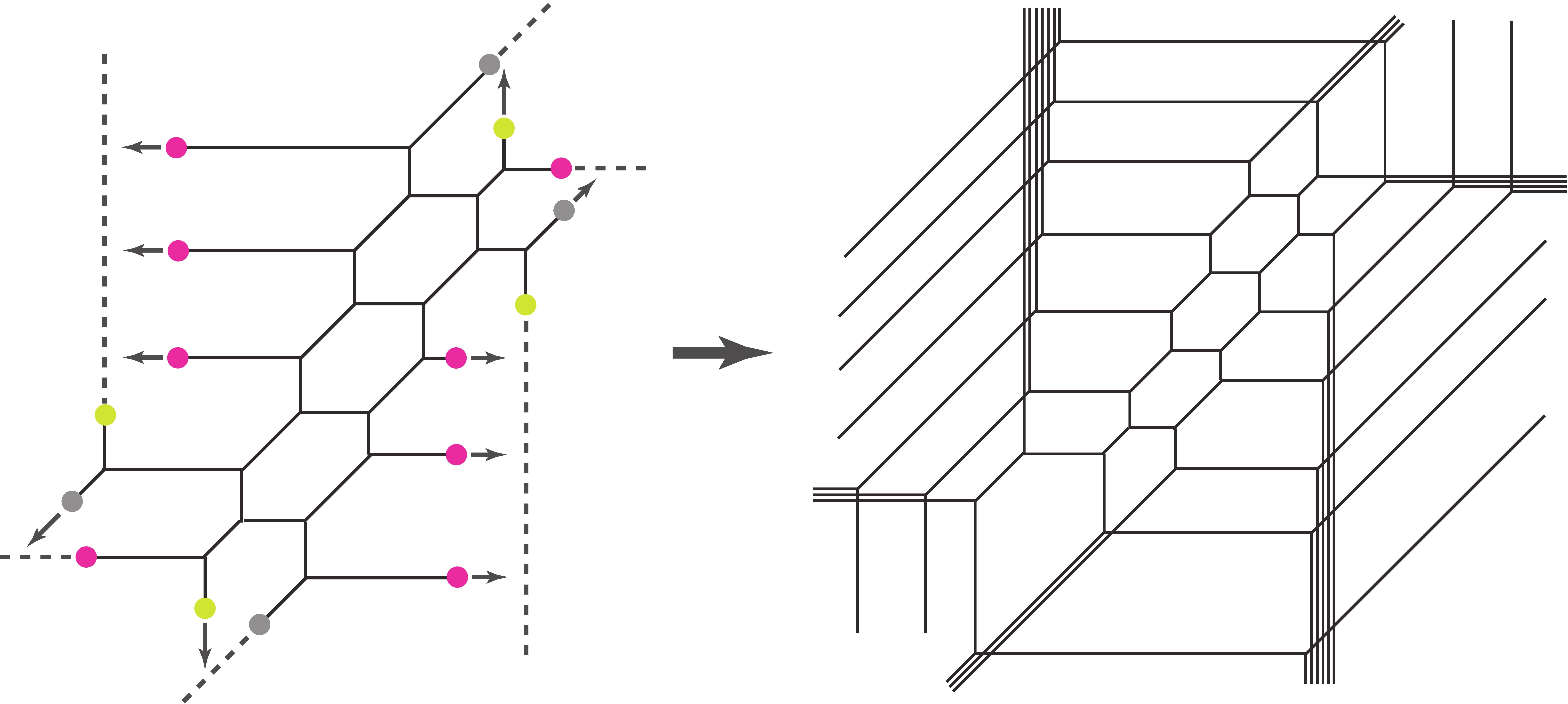}
 \end{center}
 \caption{\small A web diagram of class $\mathcal{T}$ for ``$SU(N)$ gauge theory with $2N+4$ flavors.''
 }
 \label{fig:SUN}
\end{figure}

We can also consider another higher-rank generalization of our E-string Tao web.
Based on the fact that the rank-$1$ E-string theory is the UV fixed point of 5d $SU(2)$ gauge theory with eight flavors, we can extend the gauge group to $SU(N)$ to yield a six-dimensional theory that is the UV fixed point of 5d $SU(N)$ gauge theory with a critical number of flavors. 
Figure \ref{fig:SUN} is such a web, which suggests the critical number of flavors to be ``$2N+4$.''
On the left-hand side of Figure \ref{fig:SUN}, without four 
 5-brane junctions attached to $[1,1]$ 7-branes, it is a 5-brane web for $SU(N)$ gauge theory with $2N$ flavors. With these junctions, it describes 5d $SU(N)$ gauge theory with $2N+4$ flavors and makes the spiral web shown on the right-hand side of Figure \ref{fig:SUN} as 7-branes (with arrows) across the branch cuts created by other 7-branes.
The way it makes spiral shape is as 
in Figure \ref{fig:7branemove}.
It was classified that $SU(N)$ gauge theory exists up to $2N$ flavors,
and the UV fixed point disappears if the number of hypermultiplets exceeds the limit $N_f\leq 2N$ \cite{Intriligator:1997pq}.
Our Tao-like generalization yielding to $2N+4$ flavors, on one hand, seems to be outside this classification. On the other hand,
from the existence of a consistent web diagram with spiral-direction, 
 one may expect to find a 6d fixed point in UV. \footnote{
It was conjectured in \cite{Bergman:2014kza} that the UV fixed point for 5d $SU(N)$ theory with $N_f$ flavors exists $N_f\le 2N+3$ for zero Chern-Simons level. Our observation is consistent with this conjecture.} 
Finding an F-theoretic or another realization of this 6d SCFT is an interesting direction to pursue further. 

The idea of finding a new Tao diagram is also applicable to linear quiver and 5d $T_N$ theories \cite{Gaiotto:2009we,Benini:2009gi}.
\begin{figure}[t]
 \begin{center}
\includegraphics[width=16cm, bb=0 0 3050 1288]{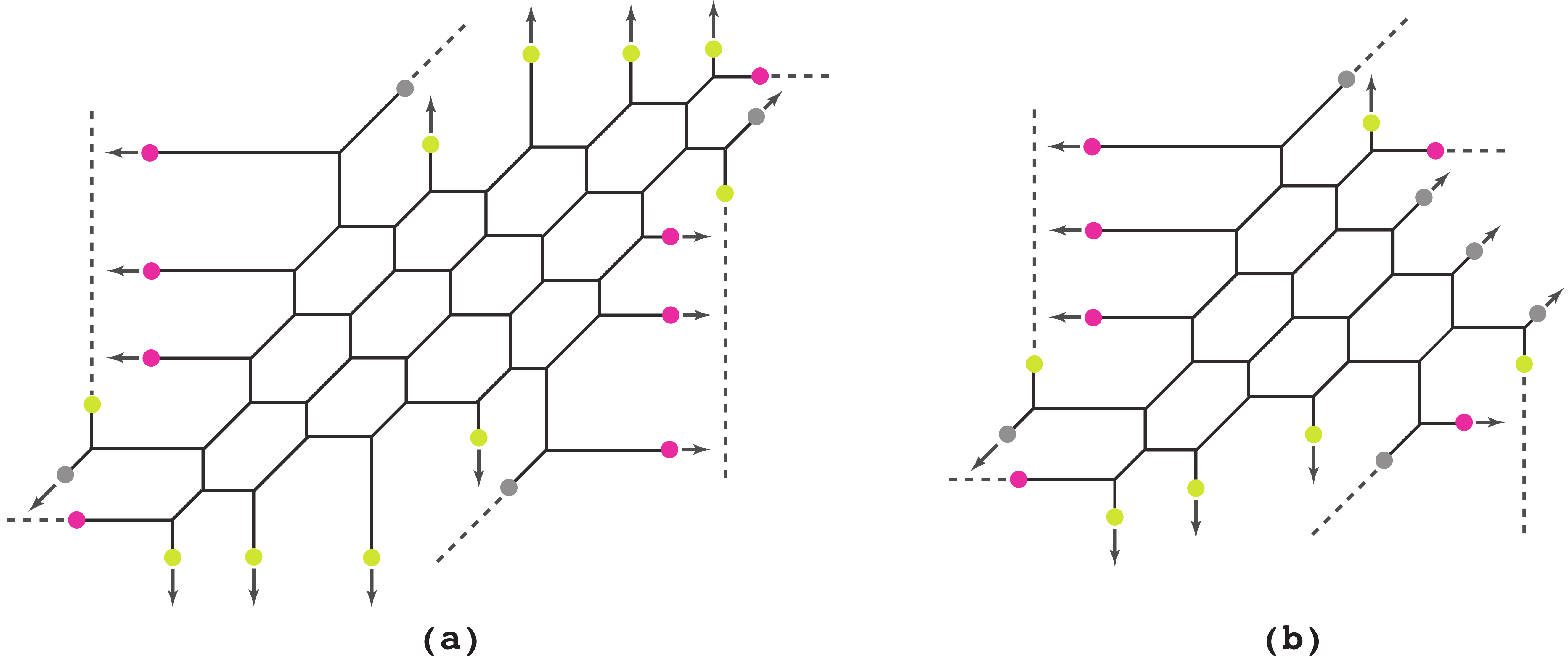}
 \end{center}
 \caption{\small Examples of web diagrams of class $\mathcal{T}$ created from linear quiver and 5d $T_N$ theories.}
 \label{fig:TNtao}
\end{figure}
Figure \ref{fig:TNtao} describes the suitably blown-up (a) linear quiver and (b) $T_N$ theory
for obtaining possible 6d SCFTs.
Figure \ref{fig:TNtao}(a) is the web of the superconformal linear quiver theory $SU(N)^M$ blown up at four points.
The gauge theory interpretation is therefore the linear quiver theory $SU(N)^M$ with an additional two pairs of fundamental hypermultiplets of $SU(N)$ at each end of the quiver.
Moving 7-branes outside, we find a consistent web diagram of class $\mathcal{T}$,
and thus there might be a 6d fixed point theory associated with this brane configuration.
Figure \ref{fig:TNtao}(b) is the $T_N$ geometry, which is $\mathbb{C}^3/\mathbb{Z}_N\times \mathbb{Z}_N$,
blown up at three points.
This is an example of class $\mathcal{T}$ created from 5d $T_N$ theory. 

One can create infinitely many webs of class $\mathcal{T}$ from conventional 5-brane webs
and identify the global symmetry of such webs 
by using the collapsing 7-branes, as was studied in \cite{DeWolfe:1999hj, Mitev:2014jza}.
Determining the global symmetry may lead to a clue to identifying the corresponding 6d theory associated with a given web. It would be interesting to see whether our construction is brane-web counterpart of the F-theoretic classification of 6d SCFTs studied recently in \cite{Heckman:2013pva}.

There are many other future directions.
For instance, the refinement $\epsilon_1+\epsilon_2\neq 0$ of our Tao partition function would be an  important direction. 
In this paper, we present a generic 5-brane configuration but the explicit computation is based on the unrefined case.
Generalization to $\epsilon_1+\epsilon_2\neq 0$ may involve further complication. %
Deriving the Seiberg--Witten curve of the E-string theory
by using Tao web is also a fruitful direction.
In \cite{Bao:2013pwa,Kim:2014nqa}, a systematic way of computing 5d Seiberg--Witten curve starting from the toric-like diagram was developed.
This method might be applicable to Tao web, leading to the Seiberg--Witten curve obtained in \cite{Eguchi:2002fc, Eguchi:2002nx}.

Considering the role of the instanton operator \cite{Lambert:2014jna,Tachikawa:2015mha}
in the context of the 5-brane web may give a hint for understanding the reason why 6d uplift occurs.
It would also be interesting to find the relation of class $\mathcal{T}$ to the followings: 
AGT correspondence for isolated SCFTs/irregular singularities 
\cite{Alday:2009aq,Wyllard:2009hg,Gaiotto:2009ma,Taki:2009zd,Kanno:2012xt,Gaiotto:2012sf,Bonelli:2011aa,Kanno:2013vi},
5d AGT correspondence \cite{Awata:2009ur,Awata:2011dc,Nieri:2013yra,Taki:2014fva}, and the corresponding $q$-Toda field theory \cite{Mitev:2014isa,Isachenkov:2014eya}.


\section*{Acknowledgements}
We would like to thank Giulio Bonelli, Tohru Eguchi, Amihay Hanany, Hee-Cheol Kim, Seok Kim, Axel Kleinschmidt, Kimyeong Lee, Kazuhiro Sakai, Yuji Tachikawa, and Piljin Yi for useful discussion and comments. 
We are grateful to the 2014 Simons Summer Workshop in Mathematics and Physics where the authors initiated the project, and the 2nd workshop on Developments in M-theory at High1. MT thanks KIAS for kind hospitality during his visit. MT is supported by the RIKEN iTHES project.

\appendix

\section{Conventions}\label{app:Convention}

\subsection{7-branes}

Given a 7-brane labeled by a pair of coprime integers $[P,Q]$, 
which is the magnetic source of dilaton--axion scalar $\tau=\chi+ie^{-\phi}$, $\tau$ 
undergoes a monodromy around the 7-brane, which is $SL(2,\mathbf{Z})$ invariant, 
and the monodromy matrix $K_{[P,Q]}$ is given by
\begin{align}
K_{[P,Q]}=
\begin{pmatrix}
1+PQ& -P^2\\
Q^2& 1-PQ
\end{pmatrix}
.
\end{align}
In our convention, a $[1,0]$ 7-brane is the familiar D7-brane.
When a $(p,q)$ 5- or $[p,q]$ 7-brane crosses counterclockwise  
a branch cut of a $[P,Q]$ 7-brane, then 5- or 7-brane experiences 
the $K_{[P,Q]}$ monodromy due to the $[P,Q]$ 7-brane, 
and the charge $(p,q)$ is altered to be another $(p',q')$ given by 
\begin{align}
\begin{pmatrix}
p'\\
q'
\end{pmatrix}=
\begin{pmatrix}
1+PQ& -P^2\\
Q^2& 1-PQ
\end{pmatrix}
\begin{pmatrix}
p\\
q
\end{pmatrix}.
\end{align}
When clockwise crossing, the brane experiences the monodromy $K^{-1}_{[P,Q]}$.

For example, consider a 7-brane associated with the flavor branes, 
which is of $[1,0]$ or $[-1, 0]$ branch cut, then the monodromy matrix is given by
\begin{align}
K_{[1,0]}=K_{[-1,0]}=
\begin{pmatrix}
1& -1\\
0& 1
\end{pmatrix}
.
\end{align}

When a $[2,1]$ brane crosses this cut clockwise, it turns into a $[3,1]$ brane, 
while a $[1,0]$ brane remains unchanged when crossing. If we have a 7-brane of 
a $[2,1]$ branch cut instead, then a $[1,0]$ brane crossing this cut 
counterclockwise turns into a $[3,1]$ brane. 

For convenience, let us use a shorthand notation for frequently appearing 7-branes: 
\begin{align}
\mathbf{A} = [1,0],\quad \mathbf{B} = [1,-1],\quad \mathbf{C} = [1, 1], \quad \mathbf{N} = [0,1].
\end{align}
It follows that 
\begin{align}\label{abca}
\mathbf{A^nBC} = \mathbf{BCA^n}, 
\end{align}
as well as 
\begin{align}\label{abnabna}
\mathbf{AB=BN},\quad \mathbf{AB=NA} \quad \mathbf{AN=CA},
\end{align}
and   
\begin{align}\label{ncbaan}
\mathbf{NC=AN},\quad \mathbf{BA=AX_{[2,-1]}} \quad \mathbf{X_{[2,-1]}N=NX_{[2,1]}}.
\end{align}

\subsection{7-brane rearrangement} 
In this paper we make full use of a special ordering of the 7-branes (\ref{Tao7branes}) associated with the $N_f=8$ theory.
Let us derive this configuration.
Consider the affine 7-brane background $\hat{E}_9$ associated with $SU(2)$ $N_f=8$ theory, namely the local $\frac{1}{2}$K3 surface
\begin{align}
\mathbf{\hat{E}_9= A^8BCBC}.
\end{align}
It follows from \eqref{abca} that 
\begin{align}
\label{A4BCA4BC}
\mathbf{A^8BCBC = A^4BCA^4BC},
\end{align}
and from \eqref{abnabna} that 
\begin{align}
\mathbf{A^4BC = A^2BN^2C = ANAN^2C = ANCANC},
\end{align}
which yields
\begin{align}
\mathbf{A^8BCBC=  (ANC)(ANC)(ANC)(ANC)}.
\end{align}
This leads to a $(p,q)$ 5-brane web with $N_f=8$: 
the $E_9$ configuration given in Figure \ref{fig:7braneE9}.
In addition, it follows from \eqref{A4BCA4BC} and \eqref{abnabna} that 
\begin{align}
\mathbf{A^8BCBC=  A^3BNCA^3BNC}.
\end{align}
Applying \eqref{ncbaan} then leads to another expression for the same 7-brane background
\begin{align}
\label{A4NXA4NX}
\mathbf{\hat{E}_9= A^3BANA^3BAN
=A^4X_{[2,-1]}NA^4X_{[2,-1]}N
=A^4NX_{[2,1]}A^4NX_{[2,1]}
}.
\end{align}

\section{Topological vertex}
The topological vertex formalism is a powerful method to compute the all-genus topological string partition functions
for the toric Calabi--Yau threefolds.
The basic building block is the vertex function
\begin{align}
C_{R_1R_2R_3}(q)
=q^{\frac{\kappa_{R_1}}{2}}
S_{R_3}(q^{\rho})
\sum_{P}
S_{R_1^t/P}(q^{R_3+\rho})
S_{R_2/P}(q^{R_3^t+\rho}),
\end{align}
where $S_R$ and $S_{R/Q}$ are the Schur and skew Schur functions.
Using  the vertex function, a topological string partition function
can be calculated as a Feynman-diagram-like quantity associated with the toric web diagram of interest \cite{Aganagic:2003db,Iqbal:2004ne}.

The convention related to the Young diagram used in the topological vertex formalism is as follows:
$\vert Y\vert:=\sum_{i}Y_i$ is the norms of $Y$, and we introduce
$\parallel Y\parallel^2:=\sum_{i}Y_i^2$.
\begin{align}
&n(Y)=\frac{\parallel Y^t\parallel^2-\vert Y\vert}{2},
\nonumber\\
&\kappa_Y=2\left( n(Y^t)-n(Y)\right).
\end{align}
$S_Y(q^{R+\rho})$
represents the Shur function $S_Y(x_1,x_2,x_3,\ldots)$ for the special arguments $x_i=q^{R_i-1+\frac{1}{2}}$.
The skew Shur function $S_{Y_1/Y_2}=\sum_{Y_3} N_{Y_2Y_3}^{Y_1}S_{Y_3}$ is defined by the fusion coefficients $N_{Y_2Y_3}^{Y_1}$:
\begin{align}
S_{Y_2}S_{Y_3}=\sum_{Y_1}N_{Y_2Y_3}^{Y_1}S_{Y_1}.
\end{align}
See \cite{MacDonnaldSymmetric} for more on the Schur functions.

\section{K\"ahler parametrization} \label{app:Kahler}

As demonstrated in Section \ref{sec:tao 5 and 7branes},
all the K\"ahler parameters can be parametrized by $y_f$, $d$ and $A$.
For the arms of the spirals, such a parametrized form is determined by solving the constraints in Section \ref{sec:tao 5 and 7branes}.
The result is as follows:
\begin{flalign}
Q_{1(1)}&=\frac{y_3}{y_2},&
Q_{2(1)}&={y_1}{y_4}, &
Q_{3(1)}&=\sqrt{\frac{y_1y_3y_5y_6}{y_2y_4y_7y_8}}\,\mathfrak{q},\\
Q_{4(1)}&=\sqrt{\frac{y_1y_3y_4y_7}{y_2y_5y_6y_8}}\,\mathfrak{q},&
Q_{5(1)}&=\sqrt{\frac{y_1y_3y_4y_5y_6y_8}{y_2y_7}}\,\mathfrak{q},&
Q_{6(1)}&=\frac{y_1}{y_8}\,\mathfrak{q}^2,
\end{flalign}
\begin{flalign}
Q_{1(2)}&={y_2}{y_4},&
Q_{2(2)}&=\sqrt{\frac{y_2y_3y_5y_6}{y_1y_4y_7y_8}}\,\mathfrak{q},&
Q_{3(2)}&=\sqrt{\frac{y_2y_3y_4y_7}{y_1y_5y_6y_8}}\,\mathfrak{q},\\
Q_{4(2)}&=\sqrt{\frac{y_2y_3y_4y_5y_6y_8}{y_1y_7}}\,\mathfrak{q},&
Q_{5(2)}&=\frac{y_2}{y_8}\,\mathfrak{q}^2,&
Q_{6(2)}&=\frac{y_3}{y_1}\,\mathfrak{q}^2,
\end{flalign}
\begin{flalign}
Q_{1(3)}&=\sqrt{\frac{y_5y_6}{y_1y_2y_3y_4y_7y_8}}\,\mathfrak{q},&
Q_{2(3)}&=\sqrt{\frac{y_4y_7}{y_1y_2y_3y_5y_6y_8}}\,\mathfrak{q},&
Q_{3(3)}&=\sqrt{\frac{y_4y_5y_6y_8}{y_1y_2y_3y_7}}\,\mathfrak{q},\\
Q_{4(3)}&=\frac{\mathfrak{q}^2}{y_3y_8},&
Q_{5(3)}&=\frac{\mathfrak{q}^2}{y_1y_2},&
Q_{6(3)}&=\frac{y_4}{y_3}\,\mathfrak{q}^2,
\end{flalign}
\begin{flalign}
Q_{1(4)}&=\frac{y_7}{y_6},&
Q_{2(4)}&={y_5}{y_8},&
Q_{3(4)}&=\sqrt{\frac{y_1y_2y_5y_7}{y_3y_4y_6y_8}}\,\mathfrak{q},\\
Q_{4(4)}&=\sqrt{\frac{y_3y_5y_7y_8}{y_1y_2y_4y_6}}\,\mathfrak{q},&
Q_{5(4)}&=\sqrt{\frac{y_1y_2y_4y_5y_7y_8}{y_3y_6}}\,\mathfrak{q},&
Q_{6(4)}&=\frac{y_5}{y_4}\,\mathfrak{q}^2,
\end{flalign}
\begin{flalign}
Q_{1(5)}&={y_6}{y_8},&
Q_{2(5)}&=\sqrt{\frac{y_1y_2y_6y_7}{y_3y_4y_5y_8}}\,\mathfrak{q},&
Q_{3(5)}&=\sqrt{\frac{y_3y_6y_7y_8}{y_1y_2y_4y_5}}\,\mathfrak{q},\\
Q_{4(5)}&=\sqrt{\frac{y_1y_2y_4y_6y_7y_8}{y_3y_5}}\,\mathfrak{q},&
Q_{5(5)}&=\frac{y_6}{y_4}\,\mathfrak{q}^2,&
Q_{6(5)}&=\frac{y_7}{y_5}\,\mathfrak{q}^2,
\end{flalign}
\begin{flalign}
Q_{1(6)}&=\sqrt{\frac{y_1y_2}{y_3y_4y_5y_6y_7y_8}}\,\mathfrak{q},&
Q_{2(6)}&=\sqrt{\frac{y_3y_8}{y_1y_2y_4y_5y_6y_7}}\,\mathfrak{q},&
Q_{3(6)}&=\sqrt{\frac{y_1y_2y_4y_8}{y_3y_5y_6y_7}}\,\mathfrak{q},\\
Q_{4(6)}&=\frac{\mathfrak{q}^2}{y_4y_7},&
Q_{5(6)}&=\frac{\mathfrak{q}^2}{y_5y_6},&
Q_{6(6)}&=\frac{y_8}{y_7}\,\mathfrak{q}^2.
\end{flalign}
Other K\"ahler parameters in the spiral are determined by
the periodic property
\begin{align}
Q_{i+6(\ell)}=\mathfrak{q}^2\cdot Q_{i(\ell)}.
\end{align}

For completeness, we list the K\"ahler parameters written in the main text:
 \begin{align*}
 Q_{mf}& = A^{-1}\,y_f  \qquad (f=1,2,3,\ldots,8), \cr
 Q_{F} &= A^{2},   \qquad Q_{B} = \frac{A^2 }{\prod^8_{f=1} y_f{}^{\frac12} } \mathfrak{q},
\end{align*}
and
\begin{align*}
\Delta_{1}&= Q_{m1}/Q_{m2}, &\Delta_{2}&= Q_{m2}Q_{m3}Q_F, & \Delta_{3}&= Q_{m4}Q_{m6}Q_B, \cr
\Delta_{4}&= Q_{m5}/Q_{m6}, &\Delta_{5}&= Q_{m6}Q_{m7}Q_F, & \Delta_{6}&= Q_{m8}Q_{m2}Q_B.
\end{align*}


\end{document}